%% file: arxiv.tex
\documentclass[USenglish,a4paper,11pt]{article}
\pdfoutput=1

\usepackage[T1]{fontenc}
\usepackage[margin=1in]{geometry}
\usepackage{microtype}
\microtypecontext{spacing=nonfrench}
\usepackage[backref=true,style=alphabetic,citestyle=alphabetic,minalphanames=3,maxalphanames=4,maxbibnames=99,maxcitenames=99]{biblatex}
\usepackage{amsmath, amsthm, amssymb}
\usepackage{graphicx}
\usepackage{mathtools}
\usepackage{enumitem}
\usepackage[dvipsnames]{xcolor} 
\usepackage{booktabs}
\usepackage{tcolorbox} 

\usepackage{hyperref}
\hypersetup{
    pdftitle={Quantum Time-Space Tradeoff for Finding Multiple Collision Pairs},
    pdfauthor={Yassine Hamoudi and Fr\'ed\'eric Magniez},
    colorlinks=true, linkcolor=BrickRed, citecolor=MidnightBlue, filecolor=blue, urlcolor=OliveGreen}
\usepackage{bookmark} 

\newcommand{\ra}{\rightarrow}
\newcommand{\la}{\leftarrow}
\newcommand{\rn}{\{0,1\}}
\newcommand{\eps}{\epsilon}
\DeclareMathOperator{\spn}{span}
\newcommand{\qub}{\ket{0}}
\newcommand{\id}{\mathbb{I}}

\providecommand\given{}
\DeclarePairedDelimiterX{\pt}[1](){\renewcommand\given{\nonscript\:\delimsize\vert\nonscript\:\mathopen{}}#1}
\DeclarePairedDelimiterX{\bc}[1][]{\renewcommand\given{\nonscript\:\delimsize\vert\nonscript\:\mathopen{}}#1}
\DeclarePairedDelimiter{\set}{\{}{\}}
\DeclarePairedDelimiter{\abs}{\lvert}{\rvert}
\DeclarePairedDelimiter{\norm}{\lVert}{\rVert}
\DeclarePairedDelimiter{\ket}{|}{\rangle}
\DeclarePairedDelimiterX{\proj}[1]{|}{|}{#1 \delimsize\rangle \delimsize\langle \mathopen{}#1}

\newcommand{\bo}{O\pt}
\newcommand{\wbo}{\widetilde{O}\pt}
\newcommand{\om}{\Omega\pt}
\newcommand{\wom}{\widetilde{\Omega}\pt}
\newcommand{\ta}{\Theta\pt}
\newcommand{\wta}{\widetilde{\Theta}\pt}

\NewDocumentCommand{\ex}{e{_}}{\mathbb{E} \IfValueT{#1}{_{#1}} \bc}
\NewDocumentCommand{\pr}{e{_}}{\Pr \IfValueT{#1}{_{#1}} \bc}
\NewDocumentCommand{\var}{e{_}}{\mathrm{Var} \IfValueT{#1}{_{#1}} \bc}

\newcommand{\ora}{\mathcal{O}}

\newcommand{\ci}{\mathcal{C}}
\newcommand{\conc}{\mathbin\Vert}

\DeclareMathOperator*{\Motimes}{\text{\raisebox{0.25ex}{\scalebox{0.8}{$\bigotimes$}}}} 
\newcommand{\rqu}{\mathsf{Q}}

\newcommand{\rf}{\mathsf{F}}

\newcommand{\rw}{\mathsf{W}}
\newcommand{\rp}{\mathsf{P}}

\newcommand{\samp}{\mathcal{S}}
\newcommand{\dis}{\mathcal{D}}
\newcommand{\unif}{\mathcal{U}}
\newcommand{\bern}{\mathcal{B}}

\newcommand{\recD}{\mathcal{R}_{\dis}}
\newcommand{\brecD}{\bar{\mathcal{R}}_{\dis}}
\newcommand{\sampD}{\samp_{\dis}}
\newcommand{\bsampD}{\mathcal{T}_{\dis}}

\newcommand{\recU}{\mathcal{R}_{\unif}}

\newcommand{\bsampU}{\mathcal{T}_{\unif}}

\newcommand{\recB}{\mathcal{R}_{\bern}}
\newcommand{\sampB}{\mathcal{S}_{\bern}}
\newcommand{\bsampB}{\mathcal{T}_{\bern}}

\newcommand{\ed}{\mathrm{ED}}
\newcommand{\suc}{\mathrm{succ}}
\newcommand{\wout}{w_\mathrm{out}}
\newcommand{\rel}{\mathrm{R}}

\theoremstyle{plain}
\newtheorem{theorem}{Theorem}[section]
\newtheorem{proposition}[theorem]{Proposition}
\newtheorem{corollary}[theorem]{Corollary}
\newtheorem{lemma}[theorem]{Lemma}
\newtheorem{fact}[theorem]{Fact}

\theoremstyle{definition}
\newtheorem{definition}[theorem]{Definition}
\newtheorem{conjecture}{Conjecture}
\newtheorem{problem}{Problem}

\newtheoremstyle{restate}{}{}{\itshape}{}{\bfseries}{.}{.5em}{\thmnote{#3}}
\theoremstyle{restate}
\newtheorem*{rtheorem}{Theorem}
\newtheorem*{rcorollary}{Corollary}
\newtheorem*{rproposition}{Proposition}

\newcommand{\algobox}[3]{
\renewcommand{\figurename}{Algorithm}
  \begin{figure}[htb] 
    \centering
    \begin{tcolorbox}
      #3
    \end{tcolorbox}
    \caption{#2}
    \label{#1}
  \end{figure}
\renewcommand{\figurename}{Figure}}

\title{Quantum Time-Space Tradeoff for \\[2mm] Finding Multiple Collision Pairs}

\author{Yassine Hamoudi\thanks{Simons Institute for the Theory of Computing, University of California, Berkeley, USA. \url{ys.hamoudi@gmail.com}}
   \and Fr\'{e}d\'{e}ric Magniez\thanks{Universit\'e Paris Cit\'e, CNRS, IRIF, F-75013, Paris, France. \url{magniez@irif.fr}}}

\date{\today}

\begin{document}

\maketitle

\begin{abstract}
  \input{Abstract.tex}
\end{abstract}

\section{Introduction}
\input{Introduction.tex}

\section{Models of computation}
\input{Model.tex}

\section{Recording query model}
\label{Sec:Recording}
\input{Record.tex}

\section{Time lower bound for Collision Pairs Finding}
\label{Sec:Coll}
\input{KColl.tex}

\section{\texorpdfstring{Time lower bound for $K$-Search}{Time lower bound for K-Search}}
  \label{Sec:QSearch}
  \input{KSearch.tex}

\section{Time-space tradeoffs}
\label{Sec:TS}
\input{Tradeoff.tex}

\section*{Acknowledgements}
  The authors want to thank the anonymous referees for their valuable comments and suggestions which helped to improve this paper. This research was supported by ANR project QUDATA ANR-18-CE47-0010, PEPR integrated project EPiQ ANR-22-PETQ-0007 part of Plan France 2030, ERC Advanced Grant PARQ, QuantERA ERA-NET Cofund project QOPT. Y.H. is also supported by the Simons Institute and DOE NQISRC QSA grant \#FP00010905.


\printbibliography[heading=bibintoc]

\end{document}

%% file: Abstract.tex
We study the problem of finding $K$ collision pairs in a random function $f : [N] \ra [N]$ by using a quantum computer. We prove that the number of queries to the function in the quantum random oracle model must increase significantly when the size of the available memory is limited. Namely, we demonstrate that any algorithm using $S$ qubits of memory must perform a number $T$ of queries that satisfies the tradeoff $T^3 S \geq \om{K^3 N}$. Classically, the same question has only been settled recently by Dinur \cite[Eurocrypt'20]{Din20c}, who showed that the Parallel Collision Search algorithm of van Oorschot and Wiener \cite{vOW99j} achieves the optimal time-space tradeoff of $T^2 S = \ta{K^2 N}$. Our result limits the extent to which quantum computing may decrease this tradeoff. Our method is based on a novel application of Zhandry's recording query technique \cite[Crypto'19]{Zha19c} for proving lower bounds in the exponentially small success probability regime. As a second application, we give a simpler proof of the time-space tradeoff $T^2 S \geq \om{N^3}$ for sorting $N$ numbers on a quantum computer, which was first obtained by Klauck, {\v{S}}palek and de Wolf~\cite{KSW07j}.

%% file: Introduction.tex
The \emph{efficiency} of a cryptographic attack is a hard-to-define concept that must express the interplay between different computational resources \cite{Wie04j,Ber05p,Ber09c}. Arguably, the two most used criteria are the \emph{time} complexity, measured for instance as the number of queries to a random oracle, and the \emph{space} complexity, which is the memory size needed to perform the attack. \emph{Time-space tradeoffs} aim at connecting these two quantities by studying how much the time increases when the available space decreases. Devising security proofs that are sensitive to memory constraints is a challenging program. Indeed, very few tools are available to quantify the extent to which the space impacts the security level of a scheme. A recent line of work \cite{TT18c,JT19c,Din20c,GJT20c} has made some progress for the case of \emph{classical} attackers with bounded memory. The development of quantum computing asks the question of whether the access to quantum operations and quantum memories may lower the security levels. The answer is unclear when taking space into account. Indeed, many quantum ``speed-ups'' come at the cost of a dramatic increase in the space requirement \cite{BHT98c,Amb07j,LZ19c}. A central open question is whether a speed-up both in term of time and space complexities is achievable for such problems?

The focus of this work is to provide time-space tradeoff lower bounds for the problem of finding \emph{multiple collision pairs} in a random function. The search for a single collision pair is one of the cornerstones of cryptanalysis. Classically, the birthday attack can be achieved by the mean of a \emph{memoryless} (i.e. logarithmic-size memory) algorithm using Pollard’s rho method \cite{Pol75j}. On the other hand, the quantum BHT algorithm \cite{BHT98c} requires fewer queries to the random function, but the product of its time and space complexities is higher than that of the classical attack! In this paper, we address the more general problem of finding \emph{multiple} collision pairs in a random function. This task plays a central role in low-memory meet-in-the-middle attacks \cite{vOW99j,Din20c}. It has applications in many problems, such as double and triple encryption \cite{vOW99j}, subset sum \cite{DDKS12c,DEM19c}, $k$-sum~\cite{Wag02c}, 3-collision~\cite{JL09c}, etc. Recently, it has also been used to attack the post-quantum cryptography candidates NTRU~\cite{vVre16j} and SIKE~\cite{ACC18c}. The celebrated classical Parallel Collision Search algorithm of van Oorschot and Wiener \cite{vOW99j} can find as many collision pairs as desired in time that depends on the available memory. The question of whether this algorithm achieves the optimal classical time-space tradeoff has been settled positively only recently by Chakrabarti and Chen \cite{CC17p} (for the case of 2-to-1 random functions) and by Dinur \cite{Din20c} (for the case of uniformly random functions). In the quantum setting, no time-space tradeoff was known prior to our work. In other words, it could have been the case that a memoryless quantum attack outperforms the Parallel Collision Search algorithm with unlimited memory capacity.

We point out that time-space tradeoffs have been studied for a long time in the complexity community \cite{BFK+81j,Bea91j,BFMadH+87j,Yao94j,BSSV03j,Abr90c,MNT93j}. The few results known in the quantum circuit model are for the Sorting problem \cite{KSW07j}, Boolean Matrix-Vector and Matrix-Matrix Multiplication \cite{KSW07j}, and Evaluating Solutions to Systems of Linear Inequalities \cite{ASW09j}. Apart from our work, all existing quantum tradeoffs are based on the hardness of Quantum Search. We use the machinery developed in our paper to give a simpler proof of the tradeoffs obtained in~\cite{KSW07j}.


\subsection{Our results}

The \emph{Collision Pairs Finding} problem asks to find a certain number $K$ of disjoint collision pairs in a random function\footnote{The notation $[m]$ for $m \in \mathbb{N}$ represents the set $\set{0,1,\dots,m-1}$.} $f : [M] \ra [N]$ where $M \geq N$. A \emph{collision pair} (or simply \emph{collision}) is a pair of values $x_1 \neq x_2$ such that $f(x_1) = f(x_2)$. Two collisions $(x_1,x_2)$ and $(x_3,x_4)$ are \emph{disjoint} if $x_1,\dots,x_4$  are all different. We define the time $T$ of an algorithm solving this problem as the number of query accesses to $f$, and the space~$S$ as the amount of memory used. We assume that the output is produced in an online fashion, meaning that a collision can be output as soon as it is discovered. The length of the output is not counted towards the space bound and the same collision may be output several times (but it contributes only once to the total count). The requirement for the collisions to be disjoint is made to simplify our proofs later on. We note that a random function $f : [N] \ra [N]$ contains $(1-2/e)N$ disjoint collisions on average \cite{FO89c}.

Classically, the single-processor Parallel Collision Search algorithm \cite{vOW99j} achieves an optimal~\cite{Din20c} time-space tradeoff of\footnote{The notation $\widetilde{\phantom{o}}$ is used to denote the presence of hidden polynomial factors in $\log(N)$ or $1/\log(N)$.} $T^2 S = \wta{K^2 N}$ for any amount of space $S$ between~$\wom{\log N}$ and $\wbo{K}$. In the quantum setting, the BHT algorithm \cite{BHT98c} can find a single collision in time $T = \wbo{N^{1/3}}$ and space $S = \wbo{N^{1/3}}$. In Algorithm~\ref{Algo:BHTvariant}, we adapt it for finding an arbitrary number $K$ of collisions at cost $T^2 S \leq \wbo{K^2 N}$. For the sake of simplicity in the analysis, we do not require these collisions to be disjoint. This is the same tradeoff as classically, except that the space parameter $S$ can hold larger values up to $\wbo{K^{2/3} N^{1/3}}$, hence the existence of a quantum speed-up when there is no memory constraint.

\begin{rproposition}[Stated in Proposition \ref{Prop:BHTvariant}]
  For any $1 \leq K \leq \bo{N}$ and $\wom{\log N} \leq S \leq \wbo{K^{2/3} N^{1/3}}$, there exists a quantum algorithm that can find $K$ collisions in a random function $f : [N] \ra [N]$ with probability at least $2/3$ by making $T = \wbo{K \sqrt{N/S}}$ queries and using $S$ qubits of memory.
\end{rproposition}

The BHT algorithm achieves the optimal time complexity for finding one collision \cite{AS04j,Zha15j}. Our first main result is to provide a similar lower bound for the problem of finding $K$ disjoint collisions. We prove that the optimal time complexity must satisfy $T \geq \om{K^{2/3} N^{1/3}}$. This bound is matched by Proposition~\ref{Prop:BHTvariant} when $S = \ta{K^{2/3} N^{1/3}}$. More precisely, we show that the optimal success probability decreases at an exponential rate in $K$ below this bound. This property is of crucial importance for proving our time-space tradeoff next. We note that, similarly to~\cite{Zha15j}, the bound is independent of the size $M$ of the domain.

\begin{rtheorem}[Stated in Theorem \ref{Thm:FindDisColl}]
  The success probability of finding $K$ disjoint collisions in a uniformly random function $f : [M] \ra [N]$ is at most $\bo{T^3/(K^2 N)}^K$ for any algorithm making $T \geq K$ quantum queries to $f$.
\end{rtheorem}

Our second main result is the next time-space tradeoff for the same problem of finding $K$ collisions in a random function. We summarize the tradeoffs known for this problem in Table~\ref{Tab:tradeoffs}. We note that the tradeoff $T^2 S \geq \om{K^2 N}$ is always stronger than $T^3 S \geq \om{K^3 N}$ since $T \geq K$.

\begin{rtheorem}[Stated in Theorem \ref{Thm:TS-DisColl}]
  Any quantum algorithm for finding $K$ disjoint collisions in a uniformly random function $f : [M] \ra [N]$ with success probability $2/3$ must satisfy a time-space tradeoff of~$T^3 S \geq \om{K^3 N}$.
\end{rtheorem}

As a simple corollary, we obtain that finding almost all collisions using a memoryless algorithm (i.e. $S = \bo{\log N}$) must require to perform at least $T \geq \om{N^{4/3}}$ quantum queries, whereas $T = N$ classical queries are clearly sufficient when there is no space restriction. We further show that any improvement to this lower bound would imply a breakthrough for the \emph{Element Distinctness} problem, which consists of finding a single collision in a random function $f : [N] \ra [N^2]$ (or, more generally, deciding if a function contains a collision). It is a long-standing open question to prove a time-space lower bound for this problem. Although there is some progress in the classical case \cite{BFMadH+87j,Yao94j,BSSV03j}, no result is known in the quantum setting. We give a reduction that converts any tradeoff for finding multiple collisions into a tradeoff for Element Distinctness. We state a particular case of our reduction below.

\begin{rcorollary}[Stated in Corollary \ref{Cor:TS-ED-All}]
  Suppose that there exists $\eps \in (0,1)$ such that any quantum algorithm for finding $\wom{N}$ disjoint collisions in a random function $f : [10N] \ra [N]$ must satisfy a time-space tradeoff of $T S^{1/3} \geq \wom{N^{4/3 + \eps}}$. Then, any quantum algorithm for solving Element Distinctness on domain size $N$ must satisfy a time-space tradeoff of $T S^{1/3} \geq \wom{N^{2/3 + 2\eps}}$.
\end{rcorollary}

We point out that $T S^{1/3} \geq \om{N^{2/3}}$ can already be deduced from the query complexity of Element Distinctness \cite{AS04j} and $S \geq 1$. We conjecture that our current tradeoff for finding $K$ collisions can be improved to $T^2 S \geq \om{K^2 N}$, which would imply $T^2 S \geq \wom{N^2}$ for Element Distinctness (Corollary~\ref{Cor:ED}). This result would be optimal~\cite{Amb07j}.

\begin{table}
  \centering\renewcommand{\arraystretch}{1.4}
  \begin{tabular}{l@{\hskip 0.2cm}c@{\hskip 0.5cm}c}
    \toprule
                   & Classical complexity                    & Quantum complexity \\ \midrule
      Upper bound: & $T^2 S \leq \wbo{K^2 N}$                & $T^2 S \leq \wbo{K^2 N}$ \\ [-5pt]
                   & \emph{when} $\wom{\log N} \leq S \leq \wbo{K}$  & \emph{when} $\wom{\log N} \leq S \leq \wbo{K^{2/3} N^{1/3}}$ \\ [-3pt]
                   & {\footnotesize Parallel Collision Search \cite{vOW99j}} & {\footnotesize Proposition~\ref{Prop:BHTvariant}} \\ \midrule
      Lower bound: & $T^2 S \geq \om{K^2 N}$                 & $T^3 S \geq \om{K^3 N}$ \\ [-3pt]
                   & {\footnotesize \cite{Din20c}}           & {\footnotesize Theorem~\ref{Thm:TS-DisColl}} \\
    \bottomrule
  \end{tabular}
\caption{Complexity to find $K$ disjoint collisions in a random function $f : [M] \ra [N]$.}
\label{Tab:tradeoffs}
\end{table}

Finally, we adapt the machinery developed in our paper to study the $K$-Search problem, which we define as the task of finding $K$ marked items (i.e. that evaluate to $1$) in a random function $f : [M] \ra \rn$ where $f(x) = 1$ with probability $p \in [0,1]$ independently for each $x$. Several variants of this problem have been considered in the literature before \cite{KSW07j,Amb10ja,S08c}, where it was shown that the success probability is exponentially small in~$K$ when the number of quantum queries is a constant fraction smaller than the complexity needed to succeed with probability~$2/3$. Our proof is the first one to consider this generic input distribution when~$K > 1$, and it is arguably simpler than previous work. The case of finding a single marked item (i.e.~$K = 1$) was solved in~\cite[Theorem 1]{HRS16c} using different techniques.

\begin{rtheorem}[Stated in Theorem \ref{Thm:QSearch}]
  Let $p \in [0,1]$. The success probability of finding $K$ marked items in a random function $f : [M] \ra \rn$ where $f(x) = 1$ with probability $p$ for each $x$ is at most~$\bo{p(T/K)^2}^K$ for any algorithm making $T \geq K$ quantum queries to $f$.
\end{rtheorem}

Note that the probability parameter must be at least $p \geq \om{K/M}$ to guarantee the existence of at least $K$ marked items in $f$ with high probability. In this regime, we can solve the $K$-Search problem with $T = \bo{K/\sqrt{p}}$ queries and success probability $2/3$ by running Grover's search on the first $\bo{K/p}$ values of $f$.

As an application of Theorem \ref{Thm:QSearch}, we reprove in Theorem~\ref{Thm:TS-DisSort} the quantum time-space tradeoff for sorting~$N$ numbers first obtained in~\cite{KSW07j}. This result requires that the circuit gates producing the output are chosen non-adaptively (we further discuss this condition in Section~\ref{Sec:Sort-TS}).

\begin{rtheorem}[Stated in Theorem~\ref{Thm:TS-DisSort}]
  Any quantum algorithm for sorting a function $f : [N] \ra \set{0,1,2}$ with success probability $2/3$ must satisfy a time-space tradeoff of $T^2 S \geq \om{N^3}$.
\end{rtheorem}


\subsection{Our techniques}

\paragraph*{Recording Query Technique.}
We use the recording query framework of Zhandry \cite{Zha19c} to upper bound the success probability of a query-bounded algorithm in finding $K$ collision pairs. This method intends to reproduce the classical strategy where the queries made by an algorithm (the \emph{attacker}) are recorded and answered with on-the-fly simulation of the oracle. Zhandry brought this technique to the quantum random oracle model by showing that, for the uniform input distribution, one can record \emph{in superposition} the queries made by a quantum algorithm. Our first technical contribution (Section~\ref{Sec:Recording}) is to simplify the analysis of Zhandry's technique and, as a byproduct, to generalize it to any product distribution on the input. We notice that there has been other independent work on extending Zhandry's recording technique \cite{HI19c,CMSZ19p,CMS19c}. Our approach does not require moving to the Fourier domain (as in \cite{CMSZ19p} for instance). Instead, it is based on defining a ``recording query operator'' that is specific to the input distribution under consideration. This operator can replace the standard quantum query operator without changing the success probability of the algorithm, but with the effect of ``recording'' the quantum queries in an additional register. We detail two recording query operators corresponding to the uniform distribution (Lemma~\ref{Lem:Coll-RecOp}) and to the product of Bernoulli distributions (Lemma~\ref{Lem:Search-RecOp}).

\paragraph*{Finding collisions with time-bounded algorithms.}
Our application of the recording technique to the Collision Pairs Finding problem has two stages. We first bound the probability that the algorithm has forced the recording of many collisions after $T$ queries. Namely, we show that the norm of the quantum state that records a new collision at the $t$-th query is on the order of $\sqrt{t/N}$ (Proposition~\ref{Prop:Coll-Prog}). This is related to the probability that a new random value collides with one of the at most $t$ previously recorded queries. The reason why the collisions have to be disjoint is to avoid the recording of more than one new collision in one query. By solving a simple recurrence relation, one gets that the amplitude of the basis states that have recorded at least~$K$ collisions after~$T$ queries is at most $\bo{T^3/(K^2 N)}^{K/2}$. We note that Liu and Zhandry \cite[Theorem 5]{LZ19c} carried out a similar analysis for the multi-collision finding problem, where they obtained the same type of bound. The second stage of our proof relates the probability of having recorded many collisions to the actual success probability of the algorithm. If we used previous approaches (notably \cite[Lemma 5]{Zha19c}), this step would degrade the upper bound on the success probability by adding a term that is polynomial in $K/N$. We preserve the exponentially small dependence on~$K$ by doing a more careful analysis of the relation between the recording and the standard query models (Proposition~\ref{Prop:Coll-Succ}). We adopt a similar approach for analyzing the $K$-Search problem in Section~\ref{Sec:QSearch}.

\paragraph*{Finding collisions with time-space bounded algorithms.}
We convert the above time-only bound into a time-space tradeoff by using the time-segmentation method \cite{BFK+81j,KSW07j}. Given a quantum circuit that solves the Collision Pairs Finding problem in time~$T$ and space~$S$, we slice it into $T/(S^{2/3} N^{1/3})$ consecutive subcircuits, each of them using $S^{2/3} N^{1/3}$ queries. If no slice can output more than $\om{S}$ collisions with high probability then there must be at least~$\om{K/S}$ slices in total, thus proving the desired tradeoff. Our above lower bound implies that it is impossible to find $\om{S}$ collisions with probability larger than $4^{-S}$ in time $S^{2/3} N^{1/3}$. We must take into account that the initial memory at the beginning of each slice carries out information from previous stages. As in previous work~\cite{Aar05j,KSW07j}, we can ``eliminate'' this memory by replacing it with the completely mixed state while decreasing the success probability by a factor of~$2^{-S}$. Thus, if a slice outputs $\om{S}$ collisions then it can be used to contradict the lower bound proved before.

\paragraph*{Element Distinctness.}
We connect the Collision Pairs Finding and Element Distinctness problems  by showing how to transform a low-space algorithm for the latter into one for the former (Proposition~\ref{Prop:reduction}). If there exists a time-$\bar{T}$ space-$\bar{S}$ algorithm for Element Distinctness on domain size $\sqrt{N}$ then we can find $\wom{N}$ collisions in a random function $f : [N] \ra [N]$ by repeatedly sampling a subset $H \subset [N]$ of size $\sqrt{N}$ and using that algorithm on the function~$f$ restricted to the domain~$H$. Among other things, we must ensure that the same collision does not occur many times and that storing the set~$H$ does not use too much memory (it turns out that 4-wise independence is sufficient for our purpose). We end up with an algorithm with time $T = \bo{N \bar{T}}$ and space $S = \bo{\bar{S}}$ for finding $\wom{N}$ collisions. Consequently, if the Element Distinctness problem on domain size~$\sqrt{N}$ can be solved with a time-space tradeoff of $\bar{T} \bar{S}^{1/3} \leq \bo{N^{1/3 + \eps}}$, then there is an algorithm for finding~$\wom{N}$ collisions that satisfies a time-space tradeoff of~$T S^{1/3} \leq \bo{N^{4/3 + \eps}}$.

%% file: Model.tex
We first present the standard model of quantum query complexity in Section~\ref{Sec:Standard}. This model is used for investigating the \emph{time complexity} of the Collision Pairs Finding problem in Section~\ref{Sec:Coll}, and of the $K$-Search problem in Section~\ref{Sec:QSearch}. Then, we describe the more general circuit model that also captures the \emph{space complexity} in Section~\ref{Sec:ModelSpace}. It is used in Section~\ref{Sec:TS} for studying time-space tradeoffs.


\subsection{Query model}
\label{Sec:Standard}

The (standard) model of quantum query complexity \cite{BdW02j} measures the number of quantum queries an algorithm (also called an ``attacker'') needs to make on an input function $f : [M] \ra [N]$ to find an output $\wout$ satisfying some predetermined relation $\rel(f,\wout)$. We present this model in more detail below.

\vspace*{10pt}
{\bf Quantum Query Algorithm.}
A $T$-query algorithm is specified by a sequence $U_0, \dots, U_T$ of unitary transformations acting on a same memory. The state $\ket{\psi}$ of that memory is made of three registers $\rqu, \rp, \rw$ where the \emph{query register} $\rqu$ holds $x \in [M]$, the \emph{phase register}~$\rp$ holds $u \in [N]$ and the \emph{working register}~$\rw$ holds some value~$w$. We represent a basis state in the corresponding Hilbert space as $\ket{x,u,w}_{\rqu\rp\rw}$. We may drop the subscript~$\rqu\rp\rw$ when it is clear from the context.
The state $\ket{\psi^f_t}$ of the algorithm after $t \leq T$ queries to some input function~$f : [M] \ra [N]$ is
  \[\ket{\psi^f_t} = U_t \ora_f U_{t-1} \cdots U_1 \ora_f U_0 \qub\]
where the oracle $\ora_f$ is defined by
  \[\ora_f \ket{x,u,w} = \omega_N^{u f(x)} \ket{x,u,w} \quad \mbox{and} \quad \omega_N = e^{\frac{2{\bf i}\pi}{N}}.\]
Note that the value of $f(x)$ is returned in the phase rather than in a state $\ket{x,u + f(x) \bmod N,w}$ as it is sometimes defined in the literature. The two kinds of queries are equivalent up to a unitary transformation but the phase encoding is more convenient to use with our framework.

The \emph{output} of the algorithm is written on some dedicated substring $\wout$ of $w$. The \emph{success probability}~$\sigma_f$ of the algorithm on input~$f$ is the probability that the output value~$\wout$ obtained by measuring the working register of $\ket{\psi^f_T}$ in the computational basis satisfies the relation $\rel(f,\wout)$. In other words, if we let $\Pi^f_{\suc}$ be the projector whose support consists of all basis states $\ket{x,u,w}$ such that the output substring $\wout$ of $w$ satisfies $\rel(f,\wout)$, then~$\sigma_f = \norm[\big]{\Pi^f_{\suc} \ket{\psi^f_T}}^2$.

\vspace*{10pt}
{\bf Oracle Register.}
Here, we describe the variant used in the adversary method \cite{Amb02j} and in Zhandry's work \cite{Zha19c}. It is represented as an interaction between an \emph{algorithm} that aims at finding a correct output $\wout$, and a superposition of \emph{oracle} inputs that respond to the queries from the algorithm.

The memory of the oracle is made of an \emph{oracle register} $\rf$ holding the description of a function $f : [M] \ra [N]$. This register is divided into $M$ subregisters $\rf_1, \dots, \rf_M$ where $\rf_x$ holds $f(x) \in [N]$ for each $x \in [M]$. The basis states in the corresponding Hilbert space are denoted by~$\ket{f}_{\rf} := \Motimes_{x \in [M]} \ket{f(x)}_{\rf_x}$. Given an input distribution $\dis$ on the set of functions~$[N]^M$, the \emph{initial state} of the oracle register is defined as $\ket{\mathcal{\dis}}_{\rf} := \sum_{f \in [N]^M} \sqrt{\pr{f \la \dis}} \ket{f}$.

The \emph{query operator} $\ora$ is a unitary transformation acting on the memory of the algorithm and the oracle. Its action is defined on each basis state by
  \[\ora \ket{x,u,w}\ket{f} = (\ora_f \ket{x,u,w})\ket{f}.\]

The joint state $\ket{\psi_t}$ of the algorithm and the oracle after $t$ queries is equal to $\ket{\psi_t} = U_t \ora U_{t-1} \cdots \allowbreak U_1 \ora U_0 (\qub\ket{\dis}) = \sum_{f \in [N]^M} \sqrt{\pr{f \la \dis}} \ket{\psi^f_t}\ket{f}$, where the unitaries $U_i$ have been extended to act as the identity on $\rf$. The \emph{success probability} $\sigma$ of a quantum algorithm on an input distribution~$\dis$ is the probability that the output value~$\wout$ and the input~$f$ obtained by measuring the working and oracle registers of the final state $\ket{\psi_T}$ satisfy the relation $\rel(f,\wout)$. In other words, if we let~$\Pi_{\suc}$ be the projector whose support consists of all basis states $\ket{x,u,w}\ket{f}$ such that the output substring $\wout$ of $w$ satisfies $\rel(f,\wout)$, then~$\sigma = \norm*{\Pi_{\suc} \ket{\psi_T}}^2$.


\subsection{Space-bounded model}
\label{Sec:ModelSpace}

We use the quantum circuit model augmented with the oracle gates of the query model defined in the above section. The \emph{time complexity}, denoted by $T$, is defined as the number of oracle gates in the circuit (which is a lower bound on the total gate complexity). The \emph{space complexity}, denoted by $S$, is the number of qubits on which the circuit is operating. The result of the computation is written on some dedicated \emph{write-only} output register $\ket{\wout}$ whose size is \emph{not counted towards the space bound}. This allows us to consider problems for which the solutions may be larger than the space bound. Formally, the output register is initially filled with some special character $\ket{\wout} = \ket{\$\$\dots\$}$ and the algorithm has access to a Boolean flag register $\ket{b}$. The only operation permitted on $\ket{\wout}$ is to copy a piece of the output (e.g.\ a collision pair) from the working space of the algorithm to the first locations holding $\$$ \emph{conditioned} on $b = 1$. This model is more general than the one described in~\cite{KSW07j,ASW09j} since the flag register allows the algorithm to choose when the output is effectively updated. We will nevertheless have to revert to the older model when analyzing the Sorting problem in Section~\ref{Sec:Sort-TS}.

We notice that, by the deferred measurement principle, any space-bounded computation that uses $T$ queries can be transformed into a $T$-query unitary algorithm as defined in Section~\ref{Sec:Standard}. Thus, any lower bound on the query complexity of a problem is also a lower bound on the time complexity of that problem in the space-bounded model. This explains our use of the query model in Section~\ref{Sec:Coll} and Section~\ref{Sec:QSearch}.

%% file: Record.tex
The quantum recording query model is a modification of the standard query model defined in Section~\ref{Sec:Standard} that is unnoticeable by the algorithm, but that allows us to track more easily the progress made towards solving the problem under consideration. The original recording model was formulated by Zhandry in \cite{Zha19c}. Here, we propose a simplified and more general version of this framework that only requires the initial state of the oracle $\ket{\dis}_{\rf}$ to be a product state $\Motimes_{x \in [M]} \ket{\dis_x}_{\rf_x}$ (instead of the uniform distribution over all basis states as in \cite{Zha19c}).

\vspace*{10pt}
{\bf Construction.}
The range $[N]$ is augmented with a new symbol $\bot$. The oracle register~$\rf$ can now contain $f : [M] \ra [N] \cup \set{\bot}$, where $f(x) = \bot$ represents the absence of knowledge from the algorithm about the image of $x$. Unlike in the standard query model, the initial state of the oracle register is independent of the input distribution and is fixed to be $\ket{\bot^M}_{\rf}$ (which represents the fact that the algorithm knows nothing about the oracle initially). We extend the query operator~$\ora$ defined in the standard query model by setting
  \[\ora \ket{x,u,w}\ket{f} = \ket{x,u,w}\ket{f} \quad \text{when $f(x) = \bot$.}\]
We take the convention that any state $\ket{x,u,w}\ket{f}$ containing $\bot$ is outside the support of~$\Pi_{\suc}$.

Given a product distribution $\dis = \dis_1 \otimes \cdots \otimes \dis_M$ on the set $[N]^M$, the initial state of the oracle register in the standard query model can be decomposed as the product state $\ket{\dis}_{\rf} = \Motimes_{x \in [M]} \ket{\dis_x}_{\rf_x}$ where $\ket{\dis_x}_{\rf_x} := \sum_{y \in [N]} \sqrt{\pr{y \la \dis_x}} \ket{y}_{\rf_x}$. The ``recording query operator''~$\recD$ is defined with respect to a family $(\samp_x)_{x\in [M]}$ of unitary operators satisfying $\samp_x \ket{\bot}_{\rf_x} = \ket{\dis_x}_{\rf_x}$ for all $x$ as follows.

\begin{definition}
  \label{Def:RecOp}
  Given a product distribution $\dis = \dis_1 \otimes \cdots \otimes \dis_M$ on the set $[N]^M$, define for each $x \in [M]$ the unitary operator $\samp_x$ acting on the register $\rf_x$ such that,
    \[
    \samp_x :
    \left\{
        \begin{array}{lll}
            \ket{\bot} & \mapsto & \ket{\dis_x}  \\[3pt]
            \ket{\dis_x} & \mapsto &\ket{\bot} \\[3pt]
            \ket{\varphi} & \mapsto &\ket{\varphi} \quad \mbox{if $\ket{\varphi} \notin \spn\set{\ket{\bot},\ket{\dis_x}}$}
        \end{array}
    \right.
    \]
  where $\ket{\dis_x} := \sum_{y \in [N]} \sqrt{\pr{y \la \dis_x}} \ket{y}$. Define $\bsampD$, $\sampD$ and the \emph{recording query operator} $\recD$ acting on all the registers $\rqu\rp\rw\rf$ such that,
   \[\bsampD = \id_{\rqu\rp\rw} \otimes \Motimes_{x \in [M]} \samp_x, \quad \sampD = \sum_{x \in [M]} \proj{x}_{\rqu} \otimes \id_{\rp\rw\rf_1 \dots \rf_{x-1}} \otimes \samp_x \otimes \id_{\rf_{x+1} \dots \rf_M}, \quad \recD = \sampD^\dagger \ora \sampD.\]
\end{definition}

Later in the paper, we study the recording query operators $\recU$ and $\recB$ related to the uniform distribution (Lemma~\ref{Lem:Coll-RecOp}) and to the product of Bernoulli distributions (Lemma~\ref{Lem:Search-RecOp}).

\vspace*{10pt}
{\bf Indistinguishability.}
The joint state of the algorithm and the oracle after $t$ queries in the recording query model is defined as
  $\ket{\phi_t} = U_t \recD U_{t-1} \cdots U_1 \recD U_0 \bigl(\qub \ket{\bot^M}\bigr)$.
Notice that the query operator~$\recD$ can only change the value of $f(x')$ (contained in the register $\rf_{x'}$) when it is applied to a state $\ket{x,u,w} \ket{f}$ such that $x = x'$. As a result, we have the following simple fact.

\begin{fact}
  \label{Fact:nonBot}
  The state $\ket{\phi_t}$ is a linear combination of basis states $\ket{x,u,w} \ket{f}$ where $f$ contains at most $t$ entries different from $\bot$.
\end{fact}

The entries of $f$ that are different from $\bot$ represent what the oracle has learned (or ``recorded'') from the queries made by the algorithm so far. In the next theorem, we show that $\ket{\phi_t}$ is related to~$\ket{\psi_t}$ (defined in Section~\ref{Sec:Standard}) by $\ket{\psi_t} = \bsampD \ket{\phi_t}$. In particular, the states $\ket{\psi_t}$ and $\ket{\phi_t}$ cannot be distinguished by the algorithm since the reduced states on the algorithm registers are identical.

\begin{theorem}
  \label{Thm:RecModel}
  Let $\dis$ be a product distribution and $(U_0, \dots, U_T)$ be a $T$-query quantum algorithm. Then, the states $\ket{\psi_T} = U_T \ora \cdots \allowbreak U_1 \ora U_0 \bigl(\qub \ket{\dis}\bigr)$ and $\ket{\phi_T} = U_T \recD \cdots U_1 \recD U_0 \bigl(\qub \ket{\bot^M}\bigr)$ obtained in the standard and recording query models respectively satisfy $\ket{\psi_T} = \bsampD \ket{\phi_T}$.
\end{theorem}

\begin{proof}
  We start by introducing the intermediate operator $\brecD = \bsampD^\dagger \ora \bsampD$. Observe that for any basis state $\ket{x,u,w} \ket{f}$ the operators $\brecD$ and $\recD$ act the same way on the registers $\rqu\rp\rf_x$ and they do not depend on the other registers. Thus, we have $\brecD = \recD$. We further observe that $U_t$ and $\bsampD$ commute for all $t$ since they depend on disjoint registers. Consequently, we have that
  \begin{align*}
    \ket{\psi_T}
      & = U_T \ora U_{T-1} \ora \cdots U_1 \ora U_0 \cdot \bsampD \pt[\big]{\qub \ket{\bot^M}} \tag*{since $\bsampD \pt[\big]{\qub \ket{\bot^M}} = \qub \ket{\dis}$}\\
      & = \bsampD \bsampD^\dagger U_T \ora \cdot \bsampD \bsampD^\dagger U_{T-1} \ora \cdots \bsampD \bsampD^\dagger U_1 \ora \cdot \bsampD \bsampD^\dagger U_0 \cdot \bsampD \pt[\big]{\qub \ket{\bot^M}} \tag*{since $\bsampD \bsampD^\dagger = I$} \\
      & = \bsampD U_T \bsampD^\dagger \cdot \ora \cdot  \bsampD U_{T-1} \bsampD^\dagger \cdot \ora \cdots \bsampD U_1 \bsampD^\dagger \cdot \ora \cdot \bsampD U_0 \pt[\big]{\qub \ket{\bot^M}} \tag*{by commutation}\\
      & = \bsampD U_T \brecD U_{T-1} \cdots U_1 \brecD U_0 \pt[\big]{\qub \ket{\bot^M}} \tag*{by definition of $\brecD$} \\
      & = \bsampD U_T \recD U_{T-1} \cdots U_1 \recD U_0 \pt[\big]{\qub \ket{\bot^M}} \tag*{since $\brecD = \recD$} \\
      & = \bsampD \ket{\phi_T} \tag*{by definition of $\ket{\phi_T}$.}
  \end{align*}
\end{proof}

%% file: KColl.tex
In this section, we upper bound the success probability of finding $K$ disjoint collisions in the query-bounded model of Section~\ref{Sec:Standard}. The problem is formally defined below. We place no restriction on the number or the repetition of collision pairs in the solution, except that at least~$K$ of them must be disjoint. This will be relevant when using the space-bounded model since the algorithm needs not remember which collision pairs have been output. For simplicity in the proof, we also assume that the solution contains the image of each collision pair under $f$.

\begin{problem}
  Let $\unif$ denote the uniform distribution over $[N]^M$. Given an integer $K$ and a function $f \sim \unif$, the \emph{Collision Pairs Finding} problem is to find a list of triples $(x_1,x_2,y_1),\dots, \allowbreak(x_{2L-1},x_{2L},y_L) \in [M]^2 \times [N]$ for some $L \geq K$ such that:
  \begin{itemize}
    \item (Collision pairs) $x_{2i-1} \neq x_{2i}$ and $f(x_{2i-1}) = f(x_{2i}) = y_i$ for all $1 \leq i \leq L$.
    \item (Mutually disjoint) There exists $K$ indices $\set{i_1,\dots,i_K} \subseteq \set{1,\dots,L}$ such that the sets $\set{x_{2i_1-1},x_{2i_1}},\set{x_{2i_2-1},x_{2i_2}},\dots,\set{x_{2i_K-1},x_{2i_K}}$ are mutually disjoint.
  \end{itemize}
\end{problem}

The proof uses the recording query model of Section~\ref{Sec:Recording}. We first describe in Section~\ref{Sec:Coll-Rec} the recording query framework associated with our input distribution. In Section~\ref{Sec:Coll-Prog}, we study the probability that an algorithm has recorded at least~$k$ collisions after~$t$ queries. We prove by induction on~$t$ and~$k$ that this quantity is exponentially small in $k$ when $t \leq \bo{k^{2/3} N^{1/3}}$ (Proposition~\ref{Prop:Coll-Prog}). Finally, in Section~\ref{Sec:Coll-Succ}, we relate this progress measure to the actual success probability (Proposition~\ref{Prop:Coll-Succ}), and we conclude that the latter quantity is exponentially small in~$K$ after $T \leq \bo{K^{2/3} N^{1/3}}$ queries (Theorem~\ref{Thm:FindDisColl}).


\subsection{Recording query operator}
\label{Sec:Coll-Rec}

The next lemma gives an alternative definition of the recording operator~$\recU$, associated with the uniform distribution $\unif$, in the standard basis. This expression will be more convenient to use later on in the proof.

\begin{lemma}
  \label{Lem:Coll-RecOp}
  If the recording query operator $\recU$ is applied to a basis state $\ket{x,u,w}\ket{f}$ where $u \neq 0$ then the register $\ket{f(x)}_{\rf_x}$ is mapped to
  \begin{align*}
    \cdot\ & \sum_{y \in [N]} \frac{\omega_N^{uy}}{\sqrt{N}} \ket{y} & \ \mbox{if $f(x) = \bot$} \ \,\\
    \cdot\ & \frac{\omega_N^{uf(x)}}{\sqrt{N}} \ket{\bot} + \frac{1 + \omega_N^{uf(x)}(N-2)}{N} \ket{f(x)} + \sum_{y \in [N] \setminus \set{f(x)}} \frac{1 - \omega_N^{uy} - \omega_N^{uf(x)}}{N} \ket{y} & \ \mbox{if $f(x) \in [N]$}
  \end{align*}
  and the other registers are unchanged. If $u = 0$ then none of the registers are changed.
\end{lemma}

\begin{proof}
  We assume that $u \neq 0$ as the query operator $\ora$ acts as the identity otherwise.
  It will be more convenient to work in the Fourier basis defined by the vectors $\ket{\widehat{v}} := \sum_{y \in [N]}  \frac{\omega_N^{vy}}{\sqrt{N}} \ket{y}$ where~$v \in [N]$. By Definition~\ref{Def:RecOp}, the action of $\recU$ on the register $\rf_x$ is:
  \begin{itemize}[leftmargin=22pt]
    \item If $f(x) = \bot$ then
        $\ket{f(x)}
          \xmapsto{\samp_x} \ket{\widehat{0}}
          \xmapsto{\ora} \ket{\widehat{u}}
          \xmapsto{\samp_x} \ket{\widehat{u}}$.
    \item If $f(x) \in [N]$ then
        $\ket{f(x)}
          \xmapsto{\samp_x} \frac{1}{\sqrt{N}} \ket{\bot} +  \allowbreak \sum\limits_{v \neq 0} \frac{\omega_N^{-vf(x)}}{\sqrt{N}} \ket{\widehat{v}}
          \xmapsto{\ora} \frac{1}{\sqrt{N}} \ket{\bot} + \sum\limits_{v \neq 0} \frac{\omega_N^{-vf(x)}}{\sqrt{N}} \ket{\widehat{u+v}}
          \xmapsto{\samp_x} \frac{1}{\sqrt{N}} \ket{\widehat{0}} + \frac{\omega_N^{uf(x)}}{\sqrt{N}} \ket{\bot} +  \sum\limits_{v \neq 0,u} \frac{\omega_N^{-vf(x)}}{\sqrt{N}} \ket{\widehat{u+v}}.$
  \end{itemize}
  One can finally check that $\sum\limits_{v \neq 0,u} \frac{\omega_N^{-vf(x)}}{\sqrt{N}} \ket{\widehat{u+v}} = \frac{\omega_N^{uf(x)}(N-2)}{N} \ket{f(x)} - \sum_{y \in [N] \setminus \set{f(x)}} \frac{\omega_N^{uy} + \omega_N^{uf(x)}}{N} \ket{y}$.
\end{proof}

The unitary $\recU$ is close to the mapping $\ket{\bot}_{\rf_x} \mapsto \sum_{y \in [N]} \frac{\omega_N^{uy}}{\sqrt{N}} \ket{y}$ and $\ket{f(x)}_{\rf_x} \mapsto \frac{\omega_N^{uf(x)}}{\sqrt{N}} \ket{\bot} + \omega_N^{uf(x)} \ket{f(x)}$ (if $f(x) \neq \bot$) up to lower-order terms of amplitude~$\bo{1/N}$. This is analogous to a ``lazy'' oracle that would choose the value of $f(x)$ uniformly at random the first time it is queried.


\subsection{Analysis of the recording progress}
\label{Sec:Coll-Prog}

We define a measure of progress based on the number of disjoint collisions contained in the oracle register of the recording query model. We first give some projectors related to this quantity.

\begin{definition}
  \label{Def:Coll-Proj}
  We define the following projectors by giving the basis states on which they project:
    \begin{itemize}
      \item $\Pi_{= k}$ and $\Pi_{\geq k}$: all basis states $\ket{x,u,w} \ket{f}$ such that $f$ contains respectively \emph{exactly} or \emph{at least}~$k$ disjoint collisions (the entries with $\bot$ are not considered as collisions).
      \item $\Pi_{= k, y}$ for $y \in \set{\bot} \cup [N]$: all basis states $\ket{x,u,w} \ket{f}$ such that (1) $f$ contains \emph{exactly}~$k$ disjoint collisions, (2) the phase multiplier $u$ is nonzero and (3) $f(x) = y$.
    \end{itemize}
\end{definition}

We can now define the measure of progress $\Delta_{t,k}$ for $t$ queries and $k$ collisions as
  \[\Delta_{t,k} = \norm*{\Pi_{\geq k} \ket{\phi_t}}\]
where $\ket{\phi_t}$ is the state after $t$ queries in the recording query model. The main result of this section is the following bound on the growth of $\Delta_{t,k}$.

\begin{proposition}
  \label{Prop:Coll-Prog}
  For all $k \leq t$, we have that $\Delta_{t,k} \leq \binom{t}{k} \pt[\Big]{\frac{4\sqrt{t}}{\sqrt{N}}}^k$.
\end{proposition}

\begin{proof}
  First, $\Delta_{0,0} = 1$ and $\Delta_{0,k} = 0$ for all $k \geq 1$ since the initial state is $\ket{\phi_0} = \qub \ket{\bot^M}$. Then, we prove that $\Delta_{t,k}$ satisfies the following recurrence relation
    \begin{equation}
      \label{Eq:Coll-Prog}
      \Delta_{t+1,k+1} \leq \Delta_{t,k+1} + 4 \sqrt{\frac{t}{N}} \Delta_{t,k}.
    \end{equation}
  From this result, we obtain that $\Delta_{t,k} \leq \binom{t}{k} \pt[\Big]{\frac{4\sqrt{t}}{\sqrt{N}}}^k$ by a simple inductive proof. In order to prove Equation~(\ref{Eq:Coll-Prog}), we first observe that $\Delta_{t+1,k+1} = \norm*{\Pi_{\geq k+1} U_{t+1} \recU \ket{\phi_t}} = \norm*{\Pi_{\geq k+1} \recU \ket{\phi_t}}$ since the unitary $U_{t+1}$ applied by the algorithm at time $t+1$ does not modify the state of the oracle register. Then, on any basis state $\ket{x,u,w} \ket{f}$, the recording query operator $\recU$ acts as the identity on the registers $\rf_{x'}$ for $x' \neq x$. Consequently, the basis states $\ket{x,u,w} \ket{f}$ in $\ket{\phi_t}$ that may contribute to $\Delta_{t+1,k+1}$ must either already contain $k+1$ disjoint collisions in $f$, or exactly~$k$ disjoint collisions in~$f$ and $u \neq 0$. This implies that, by the triangle inequality,
      \[\Delta_{t+1,k+1} \leq \Delta_{t,k+1} + \norm*{\Pi_{\geq k+1} \recU \Pi_{= k, \bot} \ket{\phi_t}} + \sum_{y \in [N]} \norm*{\Pi_{\geq k+1} \recU \Pi_{= k, y} \ket{\phi_t}}.\]

  We first bound the term $\norm*{\Pi_{\geq k+1} \recU \Pi_{= k, \bot} \ket{\phi_t}}$. Consider any basis state $\ket{x,u,w} \ket{f}$ in the support of $\Pi_{= k, \bot} \ket{\phi_t}$. By Lemma~\ref{Lem:Coll-RecOp}, we have $\recU \ket{x,u,w} \ket{f} = \sum_{y \in [N]} \frac{\omega_N^{uy}}{\sqrt{N}} \ket{x,u,w}\ket{y}_{\rf_x} \otimes \allowbreak \mathop{\Motimes}_{x' \neq x} \ket{f(x')}_{\rf_{x'}}$. There are at most $t$ entries in $f$ that can collide with the value contained in the register~$\rf_x$ by Fact~\ref{Fact:nonBot}. Thus, we have $\norm{\Pi_{\geq k+1} \recU \ket{x,u,w} \ket{f}} \leq \sqrt{t/N}$. Since any two basis states in the support of~$\Pi_{= k, \bot}$ remain orthogonal after $\Pi_{\geq k+1} \recU$ is applied, we obtain that $\norm{\Pi_{\geq k+1} \recU \Pi_{= k, \bot} \ket{\phi_t}} \leq \sqrt{t/N} \norm{\Pi_{= k, \bot} \ket{\phi_t}} \leq \sqrt{t/N} \Delta_{t,k}$.

  We now bound the term $\norm*{\Pi_{\geq k+1} \recU \Pi_{= k, y} \ket{\phi_t}}$ for any $y \in [N]$. Consider any state~$\ket{x,u,w} \ket{f}$ in the support of $\Pi_{= k, y}\ket{\phi_t}$. By Lemma~\ref{Lem:Coll-RecOp}, we have $\recU \ket{x,u,w} \ket{f} = \ket{x,u,w} \pt[\Big]{\frac{\omega_N^{uf(x)}}{\sqrt{N}} \ket{\bot}_{\rf_x} + \frac{1 + \omega_N^{uf(x)}(N-2)}{N} \ket{f(x)}_{\rf_x} + \sum_{y' \neq f(x)} \allowbreak \frac{1 - \omega_N^{uy'} - \omega_N^{uf(x)}}{N} \ket{y'}_{\rf_x}} \otimes \allowbreak \mathop{\Motimes}_{x' \neq x} \ket{f(x')}_{\rf_{x'}}$. By Fact~\ref{Fact:nonBot}, there are at most $t$ terms in this sum that can be in the support of $\Pi_{\geq k+1}$. Thus, $\norm{\Pi_{\geq k+1} \recU \ket{x,u,w} \ket{f}} \leq 3\sqrt{t}/N$ and $\norm{\Pi_{\geq k+1} \recU \Pi_{= k, y} \ket{\phi_t}} \allowbreak \leq 3\sqrt{t}/N \norm{\Pi_{= k, y} \ket{\phi_t}}$.

  We conclude that $\Delta_{t+1,k+1} \leq \Delta_{t,k+1} + \sqrt{t/N} \Delta_{t,k} + \sum_{y \in [N]} 3\sqrt{t}/N \allowbreak \norm{\Pi_{= k, y} \ket{\phi_t}} \allowbreak \leq \Delta_{t,k+1} + \sqrt{t/N} \Delta_{t,k} + 3\sqrt{t/N} \sqrt{\sum_{y \in [N]} \norm{\Pi_{= k, y} \ket{\phi_t}}^2} \leq \Delta_{t,k+1} + \sqrt{t/N} \Delta_{t,k} + 3\sqrt{t/N} \Delta_{t,k}$, where the second step is by Cauchy-Schwarz' inequality.
\end{proof}


\subsection{From the recording progress to the success probability}
\label{Sec:Coll-Succ}

We connect the success probability $\sigma = \norm*{\Pi_{\suc} \ket{\psi_T}}^2$ in the standard query model to the final progress $\Delta_{T,k}$ in the recording query model after $T$ queries. We show that if the algorithm has made no significant progress for recording more than $k \leq K$ collisions then it needs to ``guess'' the positions of  $K - k$ other collisions. Classically, the probability to find the values of $K-k$ collisions that have not been queried would be at most $1/N^{K-k}$. Here, we show similarly that if a unit state contains at most $k$ collisions in the quantum recording model, then after mapping it to the standard query model (by applying the operator $\bsampU$ of Definition~\ref{Def:RecOp}) the probability that the output register contains the correct positions of $K$ collisions is at most $(2K+1)^2 (2 K/N)^{K-k}$.

\begin{proposition}
  \label{Prop:Coll-Succ}
  For any state $\ket{\phi}$, we have $\norm{\Pi_{\suc} \bsampU \Pi_{= k} \ket{\phi}} \leq  (2K+1) \pt[\Big]{\sqrt{\frac{2K}{N}}}^{K-k} \norm{\Pi_{= k} \ket{\phi}}$.
\end{proposition}

\begin{proof}
  We assume that the output $\wout$ of the algorithm consists of exactly $K$ collision pairs $(x_1,x_2,y_1),\dots, \allowbreak(x_{2K-1},x_{2K},y_K) \in [M]^2 \times [N]$ (if there are $L \geq K$ triples, we can copy the first $K$  disjoint collisions to a new register $\wout'$ to which the rest of the proof is applied). The output is correct if the input function $f : [M] \ra [N]$ (in the standard query model) satisfies $f(x_{2i-1}) = f(x_{2i}) = y_i$ for all $1 \leq i \leq K$, and the values $x_1,x_2,\dots,x_{2K}$ are all different. By definition, the support of $\Pi_{\suc}$ consists of all basis states $\ket{x,u,w} \ket{f}$ such that the output substring $\wout$ of $w$ satisfies these conditions.

  We define a new family of projectors $\tilde{\Pi}_{a,b}$, where $0 \leq a + b \leq 2K$, whose supports consist of all basis states $\ket{x,u,w} \ket{f}$ satisfying the following conditions:
  \begin{itemize}
    \item[(A)] The output substring $\wout$ is made of $K$ triples $(x_1,x_2,y_1),\dots, \allowbreak (x_{2K-1},x_{2K},y_K) \in [M]^2 \times [N]$ where the~$x_i$ are all different.
    \item[(B)] There are exactly $a$ indices $i \in [2K]$ such that $f(x_i) = \bot$.
    \item[(C)] There are exactly $b$ indices $i \in [2K]$ such that $f(x_i) \neq \bot$ and $f(x_i) \neq y_{\lceil i/2 \rceil}$.
  \end{itemize}

  For any state $\ket{x,u,w} \ket{f}$ in the support of $\tilde{\Pi}_{a,b}$, we claim that
    \begin{equation}
      \label{Eq:basis}
      \norm*{\Pi_{\suc} \bsampU \ket{x,u,w} \ket{f}} \leq \pt*{\frac{1}{\sqrt{N}}}^a \pt*{\frac{1}{N}}^b.
    \end{equation}
  Indeed, we have $\bsampU = \id \otimes \Motimes_{x' \in [M]} \samp_{x'}$ and by Definition~\ref{Def:RecOp} the action of $\samp_{x_i}$ on the register $\ket{f(x_i)}_{\rf_{x_i}}$ is $\ket{f(x_i)} \mapsto \frac{1}{\sqrt{N}} \sum_{y \in [N]} \ket{y}$ if $f(x_i) = \bot$, and $\ket{f(x_i)} \mapsto \frac{1}{\sqrt{N}} \ket{\bot} + (1 - \frac{1}{N}) \ket{f(x_i)} - \frac{1}{N} \sum_{y \in [N] \setminus \set{f(x_i)}} \ket{y}$ otherwise. The projector $\Pi_{\suc}$ only keeps the term $\ket{y_{\lceil i/2 \rceil}}$ in these sums, which implies Equation~(\ref{Eq:basis}).

  Let us now consider any linear combination $\ket{\varphi} = \sum_{x,u,w,f} \alpha_{x,u,w,f} \allowbreak \ket{x,u,w} \ket{f}$ of basis states that are in the support of $\tilde{\Pi}_{a,b}$. We claim that
    \begin{equation}
      \label{Eq:comb}
      \norm{\Pi_{\suc} \bsampU \ket{\varphi}} \leq \pt[\bigg]{\sqrt{\frac{2K}{N}}}^{a+b} \norm{\ket{\varphi}}.
    \end{equation}
  Given two states $\ket{x,u,w} \ket{f}$ and $\ket{\bar{x},\bar{u},\bar{w}} \ket{\bar{f}}$ where $\wout = ((x_1,x_2,y_1),\dots, (x_{2K-1}, \allowbreak x_{2K},y_K))$ is the output of $w$, if the tuples $\pt[\big]{x, u, w, (f(x'))_{x' \notin \set{x_1,\dots,x_{2K}}}}$ and $\pt[\big]{\bar{x}, \bar{u}, \bar{w}, \allowbreak (\bar{f}(x'))_{x' \notin \set{x_1,\dots,x_{2K}}}}$ are different then $\Pi_{\suc} \bsampU \ket{x,u,w} \ket{f}$ must be orthogonal to $\Pi_{\suc} \bsampU \ket{\bar{x},\bar{u},\bar{w}} \ket{\bar{f}}$. Moreover, for any $\wout =((x_1,x_2,y_1),\dots,(x_{2K-1},x_{2K},y_K))$ that satisfies condition (A), there are $\binom{2K}{a}\binom{2K-a}{b}(N-1)^b \leq (2K)^{a+b} N^b$ different ways to choose $(f(x_i))_{i \in [2K]}$ that satisfy conditions (B) and (C). Let us write $w_{\vec{x}} = \set{x_1,\dots,x_{2K}}$ when the output substring $\wout$ of $w$ contains $x_1,\dots,x_{2K}$. Then, by using the Cauchy-Schwarz inequality and Equation~(\ref{Eq:basis}), we get that
    \begin{align*}
      \norm{\Pi_{\suc} \bsampU \ket{\varphi}}^2
       & = \sum_{x, u, w, (f(x'))_{x' \notin w_{\vec{x}}}} \norm[\Big]{\sum_{(f(x'))_{x' \in w_{\vec{x}}}} \alpha_{x,u,w,f} \Pi_{\suc} \bsampU \ket{x,u,w} \ket{f}}^2 \\
       & \leq \sum_{x, u, w, (f(x'))_{x' \notin w_{\vec{x}}}} \pt[\bigg]{\sum_{(f(x'))_{x' \in w_{\vec{x}}}} \abs{\alpha_{x,u,w,f}}^2} \pt[\bigg]{\sum_{(f(x'))_{x' \in w_{\vec{x}}}} \norm{\Pi_{\suc} \bsampU \ket{x,u,w} \ket{f}}^2} \\
       & \leq \norm{\ket{\varphi}}^2 \cdot (2K)^{a+b} N^b \pt*{\frac{1}{N}}^a \pt*{\frac{1}{N^2}}^b \\
       & = \pt*{\frac{2K}{N}}^{a+b} \norm*{\ket{\varphi}}^2,
    \end{align*}
  which proves Equation~(\ref{Eq:comb}). Observe now that the support of $\Pi_{= k}$ is contained into the union of the supports of $\tilde{\Pi}_{a,b}$ for $a+b \geq K-k$, augmented with the basis states $\ket{x,u,w} \ket{f}$ that do not satisfy the condition (A) described above (these states are zeroed out when applying $\Pi_{\suc} \bsampU$). Thus, by the triangle inequality, $\norm*{\Pi_{\suc} \bsampU \Pi_{= k} \ket{\phi}} \leq \sum_{a+b \geq K-k} \norm{\Pi_{\suc} \bsampU \tilde{\Pi}_{a,b} \Pi_{= k} \ket{\phi}}$. This is at most $\sum_{a+b \geq K-k} \pt[\Big]{\sqrt{\frac{2K}{N}}}^{a+b} \norm{\tilde{\Pi}_{a,b} \Pi_{= k} \ket{\phi}}$ by Equation~(\ref{Eq:comb}). Finally, by Cauchy-Schwarz' inequality and the fact that the supports of the projectors $\tilde{\Pi}_{a,b}$ are disjoint, we have $\norm*{\Pi_{\suc} \bsampU \Pi_{= k} \ket{\phi}} \leq \sqrt{\sum_{K-k \leq a+b \leq 2K} \pt*{\frac{2K}{N}}^{a+b}} \sqrt{\sum_{a,b} \norm{\tilde{\Pi}_{a,b} \Pi_{= k} \ket{\phi}}^2} \allowbreak \leq (2K+1) \pt[\Big]{\sqrt{\frac{2K}{N}}}^{K-k} \norm*{\Pi_{= k} \ket{\phi}}$.
\end{proof}

We can now conclude the proof of the main result of this section.

\begin{theorem}
  \label{Thm:FindDisColl}
  The success probability of finding $K$ disjoint collisions in a uniformly random function $f : [M] \ra [N]$ is at most $\bo{T^3/(K^2 N)}^K$ for any algorithm making $T \geq K$ quantum queries to~$f$.
\end{theorem}

\begin{proof}
  Let $\ket{\psi_T}$ (resp. $\ket{\phi_T}$) denote the state of the algorithm after $T$ queries in the standard (resp. recording) query model. We recall that $\ket{\psi_T} = \bsampU \ket{\phi_T}$ (Theorem~\ref{Thm:RecModel}). Thus, by using the fact that $\id = \sum_{k=0}^{K-1} \Pi_{= k} + \Pi_{\geq K}$, the success probability $\sigma = \norm{\Pi_{\suc} \ket{\psi_T}}^2$ satisfies
    \begin{align*}
     \sqrt{\sigma}
      & \leq \sum_{k = 0}^{K-1} \norm{\Pi_{\suc} \bsampU \Pi_{= k} \ket{\phi_T}} + \norm{\Pi_{\suc} \bsampU \Pi_{\geq K} \ket{\phi_T}} \tag*{by the triangle inequality} \\
      & \leq \sum_{k = 0}^{K-1} (2K+1) \pt*{\frac{\sqrt{2K}}{\sqrt{N}}}^{K-k} \norm{\Pi_{= k} \ket{\phi_T}} + \norm{\Pi_{\geq K} \ket{\phi_T}} \tag*{by Proposition~\ref{Prop:Coll-Succ}}\\
      & \leq \sum_{k = 0}^{K} (2K+1) \pt*{\frac{\sqrt{2K}}{\sqrt{N}}}^{K-k} \binom{T}{k} \pt*{\frac{4\sqrt{T}}{\sqrt{N}}}^k \tag*{by Proposition~\ref{Prop:Coll-Prog}} \\
      & \leq (2K+1) \pt*{\frac{4\sqrt{T}}{\sqrt{N}}}^K \sum_{k = 0}^{K} \binom{T}{k} \tag*{since $K \leq T$}\\
      & \leq (2K+1) \pt*{\frac{4e T^{3/2}}{K\sqrt{N}}}^K \tag*{since $\sum_{k = 0}^{K} \binom{T}{k} \leq (eT/K)^K$.}
    \end{align*}
\end{proof}

%% file: KSearch.tex
In this section, we illustrate the use of the recording query model to upper bound the success probability of a query-bounded algorithm on a \emph{non-uniform} input distribution. Specifically, we consider the following variant of the $K$-Search problem.

\begin{problem}
  Let $p \in [0,1]$ and define the distribution $\bern$ over the functions $f : [M] \ra \rn$ where $f(x)=1$ with probability $p$ independently for each $x$. Given an integer $K$ and a function~$f \sim \bern$, the \emph{$K$-Search} problem is to find~$K$ distinct values $x_1,\dots,x_K \in [M]$ such that $f(x_i) = 1$ for all $i$.
\end{problem}

We show that, similarly to the classical setting where a query can reveal a marked item with probability~$p$, the \emph{amplitude} of the basis states that record a new~$1$ increases by a factor of~$\sqrt{p}$ after each query (Proposition~\ref{Prop:QSearch-Prog}). Thus, the amplitude of the basis states that have recorded at least $K$ ones after $T$ queries is at most $\bo{T\sqrt{p}/K}^K$. This implies that any algorithm with $T \ll \bo{K/\sqrt{p}}$ queries is likely to output coordinates that have not been recorded, each of which decreases the success probability by a factor of $\bo{p}$ (Proposition~\ref{Prop:QSearch-Succ}).

Although the proof consists again of bounding a certain progress measure related to the success probability, the analysis differs from that of the above section by requiring to adapt the projectors (Definition~\ref{Def:QSearch-Proj}) and the orthogonality arguments (Proposition~\ref{Prop:QSearch-Succ}) to the new input distribution and type of progress.


\subsection{Recording query operator}
\label{Sec:QSearch-Rec}

The next lemma gives an alternative definition of the recording operator~$\recB$, associated with the distribution $\bern$, in the standard basis.

\begin{lemma}
  \label{Lem:Search-RecOp}
  If the recording query operator $\recB$ is applied to a basis state $\ket{x,u,w}\ket{f}$ where $u = 1$ then the register $\ket{f(x)}_{\rf_x}$ is mapped to
    \[
        \arraycolsep=1.6pt\def\arraystretch{1.4}
        \begin{array}{lrcrcrl}
            \cdot\ & (1-2p)\,           \ket{\bot}   & + &
            2p\sqrt{1-p}\,         \ket{0}      & - &
            2\sqrt{p}(1-p)\,           \ket{1}      & \quad\quad \mbox{if $f(x) = \bot$}  \\[1mm]
            \cdot\ & 2p\sqrt{1-p}\,         \ket{\bot}   & + &
            (1 - 2p(1-p))\, \ket{0}      & + &
            2\sqrt{p}(1-p)^{3/2}\,           \ket{1}      & \quad\quad \mbox{if $f(x) = 0$}  \\[1mm]
            \cdot\ & - 2\sqrt{p}(1-p)\,        \ket{\bot}   & + &
            2\sqrt{p}(1-p)^{3/2}\,         \ket{0}      & + &
            (1 - 2(1-p)^2)\,        \ket{1}      & \quad\quad \mbox{if $f(x) = 1$}
      \end{array}
    \]
  and the other registers are unchanged. If $u = 0$ then none of the registers are changed.
\end{lemma}

\begin{proof}
  We assume that $u = 1$ as the query operator $\ora$ acts as the identity otherwise. Let $\ket{+} = \sqrt{1-p} \ket{0} + \sqrt{p} \ket{1}$ and $\ket{-} = \sqrt{p} \ket{0} - \sqrt{1-p} \ket{1}$. Define the following $3 \times 3$ matrices,
  \[M_B = \begin{pmatrix} 1 & 0 & 0 \\ 0 & \sqrt{p} & \sqrt{1-p} \\ 0 & -\sqrt{1-p} & \sqrt{p} \end{pmatrix}, \quad
    M_{\samp} = M_B^{\perp}\begin{pmatrix} 0 & 1 & 0 \\ 1 & 0 & 0 \\ 0 & 0 & 1 \end{pmatrix}M_B,\quad
    M_{\ora} = \begin{pmatrix} 1 & 0 & 0 \\ 0 & 1 & 0 \\ 0 & 0 & -1 \end{pmatrix}.\]
  The matrix $M_B$ represents the change of basis from $\{\ket{\bot},\ket{0},\ket{1}\}$ to $\{\ket{\bot},\ket{+},\ket{-}\}$. The matrix~$M_{\samp}$ corresponds to $\samp_x$ (Definition~\ref{Def:RecOp}) expressed in the $\{\ket{\bot},\ket{0},\ket{1}\}$ basis. The matrix~$M_{\ora}$ corresponds to the action of the query operator $\ora$ on the register $\rf_x$, expressed in the $\{\ket{\bot},\ket{0},\ket{1}\}$ basis, when it is applied to a state $\ket{x,u,w}\ket{f}$ with $u = 1$. The three equations in the lemma are given by the columns of the matrix product $M_{\samp} M_{\ora} M_{\samp}$ by definition of $\recB = \sampB^{\dagger} \ora \sampB$.
\end{proof}

If $p \ll 1$, the above lemma shows that $\recB$ is close to the mapping $\ket{\bot}_{\rf_x} \mapsto \ket{\bot} - 2\sqrt{p} \ket{1}$, $\ket{0}_{\rf_x} \mapsto \ket{0} + 2\sqrt{p} \ket{1}$, $\ket{1}_{\rf_x} \mapsto - \ket{1} + 2\sqrt{p}(\ket{0} - \ket{\bot})$ up to lower order terms of amplitude~$\bo{p}$. This is again similar to the behavior of a lazy oracle.


\subsection{Analysis of the recording progress}
\label{Sec:QSearch-Prog}

The measure of progress is based on the number of ones contained in the oracle register. We first give some projectors related to this quantity.

\begin{definition}
  \label{Def:QSearch-Proj}
  We define the following projectors by giving the basis states on which they project:
    \begin{itemize}
      \item $\Pi_{= k}$ and $\Pi_{\geq k}$: all basis states $\ket{x,u,w} \ket{f}$ such that $f$ contains respectively \emph{exactly} or \emph{at least} $k$ coordinates equal to $1$.
      \item $\Pi_{= k, \bot}$ and $\Pi_{= k, 0}$: all basis states $\ket{x,u,w} \ket{f}$ such that (1) $f$ contains \emph{exactly} $k$ coordinates equal to $1$, (2) the phase multiplier is $u = 1$ and (3) $f(x) = \bot$ or $f(x) = 0$ respectively.
    \end{itemize}
\end{definition}

We define the measure of progress $\Delta_{t,k} = \norm*{\Pi_{\geq k} \ket{\phi_t}}$ where $\ket{\phi_t}$ is the state after $t$ queries in the recording query model. We obtain the following bound on the growth of $\Delta_{t,k}$.

\begin{proposition}
  \label{Prop:QSearch-Prog}
  For all $k \leq t$, we have that $\Delta_{t,k} \leq \binom{t}{k} (4 \sqrt{p})^k$.
\end{proposition}

\begin{proof}
  First, $\Delta_{0,0} = 1$ and $\Delta_{0,k} = 0$ for all $k \geq 1$. Then, we prove that $\Delta_{t,k}$ satisfies the following recurrence relation
    \begin{equation}
      \label{Eq:QSearch-Prog}
      \Delta_{t+1,k+1} \leq \Delta_{t,k+1} + 4 \sqrt{p} \Delta_{t,k}.
    \end{equation}
  From this result, we obtain that $\Delta_{t,k} \leq \binom{t}{k} (4 \sqrt{p})^k$ by induction. Similarly to Proposition~\ref{Prop:Coll-Prog}, the proof of Equation~(\ref{Eq:QSearch-Prog}) uses that
    \[\Delta_{t+1,k+1} \leq \Delta_{t,k+1} + \norm*{\Pi_{\geq k+1} \recB \Pi_{= k, \bot} \ket{\phi_t}} + \norm*{\Pi_{\geq k+1} \recB \Pi_{= k, 0} \ket{\phi_t}}.\]

  We first bound the term $\norm{\Pi_{\geq k+1} \recB \Pi_{= k, \bot} \ket{\phi_t}}$. Consider any state $\ket{x,u,w} \ket{f}$ in the support of $\Pi_{= k, \bot} \ket{\phi_t}$. By Lemma~\ref{Lem:Search-RecOp}, we have $\Pi_{\geq k+1} \recB \ket{x,u,w} \ket{f} = - 2\sqrt{p}(1-p) \ket{x,u,w} \ket{1}_{\rf_x} \otimes \allowbreak \mathop{\Motimes}_{x' \neq x} \ket{f(x')}_{\rf_{x'}}$. Since any two basis states in the support of $\Pi_{= k, \bot}$ remain orthogonal after $\Pi_{\geq k+1} \recB$ is applied, we have $\norm{\Pi_{\geq k+1} \recB \Pi_{= k, \bot} \ket{\phi_t}} = 2 \sqrt{p}(1-p) \norm*{\Pi_{= k, \bot} \ket{\phi_t}} \leq 2 \sqrt{p}(1-p) \Delta_{t,k}$.

  Similarly, for $\ket{x,u,w} \ket{f}$ in the support of $\Pi_{= k, 0} \ket{\phi_t}$ we have $\norm{\Pi_{\geq k+1} \recB \ket{x,u,w} \ket{f}} = 2\sqrt{p}(1-p)^{3/2}$. Thus, $\norm{\Pi_{\geq k+1} \recB \Pi_{= k, 0} \ket{\phi_t}} = 2\sqrt{p}(1-p)^{3/2} \norm{\Pi_{= k, 0} \ket{\phi_t}} \leq 2 \sqrt{p}(1-p)^{3/2} \Delta_{t,k}$. We can now conclude the proof,
    \[
      \Delta_{t+1,k+1} \leq \Delta_{t,k+1} + 2\sqrt{p}(1-p) \Delta_{t,k} + 2 \sqrt{p}(1-p)^{3/2} \Delta_{t,k}  \leq \Delta_{t,k+1} + 4 \sqrt{p} \Delta_{t,k}.
     \]
\end{proof}


\subsection{From the recording progress to the success probability}
\label{Sec:QSearch-Succ}

We connect the success probability $\sigma = \norm*{\Pi_{\suc} \ket{\psi_T}}^2$ in the standard query model to the final progress $\Delta_{T,k}$ in the recording query model. Classically, the probability to find~$K-k$ marked items at positions that have not been queried would be $p^{K-k}$. Here, we show similarly that if a unit state contains~$k$ ones in the quantum recording model then, after mapping it to the standard query model, the probability that the output register contains the correct positions of $K$ marked items is at most~$3^K p^{K-k}$.

\begin{proposition}
  \label{Prop:QSearch-Succ}
  For any $\ket{\phi}$, we have $\norm{\Pi_{\suc} \bsampB \Pi_{= k} \ket{\phi}} \leq  3^{K/2} p^{(K-k)/2} \norm{\Pi_{= k} \ket{\phi}}$.
\end{proposition}

\begin{proof}
  Let $\ket{x,u,w} \ket{f}$ be any basis state in the support of $\Pi_{= k}$. The output value $\wout$ is a substring of $w$ made of $K$ distinct values $x_1, \dots, x_K \in [M]$ indicating positions where the input~$f$ is supposed to evaluate to $1$. By definition of $\Pi_{= k}$, we have $f(x_i) \neq 1$ for at least $K - k$ indices $i \in [K]$. For each such index $i$, after applying $\bsampB = \id \otimes \Motimes_{x' \in [M]} \samp_{x'}$, the amplitude of $\ket{1}_{\rf_{x_i}}$ is~$\sqrt{p}$ (if $f(x_i) = \bot$) or $-\sqrt{p(1-p)}$ (if $f(x_i) = 0$) by Definition~\ref{Def:RecOp}. Consequently,
    \begin{equation}
      \label{Eq:searchDec}
      \norm*{\Pi_{\suc} \bsampB \ket{x,u,w} \ket{f}} \leq p^{(K-k)/2}.
    \end{equation}

  We define $w_{\vec{x}} = \set{x_1,\dots,x_{K}}$ when the output substring $\wout$ of $w$ contains $x_1,\dots,x_{K}$. For any two basis states $\ket{x,u,w} \ket{f}$ and $\ket{\bar{x}, \bar{u}, \bar{w}} \ket{\bar{f}}$, if
    $\pt[\big]{x, u, w, (f(x'))_{x' \notin w_{\vec{x}}}} \neq \pt[\big]{\bar{x}, \bar{u}, \bar{w}, (\bar{f}(x'))_{x' \notin w_{\vec{x}}}}$
  then $\Pi_{\suc} \bsampB \ket{x,u,w} \ket{f}$ is orthogonal to $\Pi_{\suc} \bsampB \ket{\bar{x}, \bar{u}, \bar{w}} \ket{\bar{f}}$. There are $3^K$ choices for $\ket{x,u,w} \ket{f}$ once we set the value of $\pt{x, u, w, (f(x'))_{x' \notin w_{\vec{x}}}}$ since it remains to choose $f(x') \in \set{\bot,0,1}$ for~$x' \in w_{\vec{x}}$. Consider now any state $\ket{\phi}$ and denote $\ket{\varphi} = \Pi_{= k} \ket{\phi} = \sum_{x,u,w,f} \alpha_{x,u,w,f} \allowbreak \ket{x,u,w} \ket{f}$. By using the Cauchy--Schwarz inequality and Equation~(\ref{Eq:searchDec}), we get that
  \begin{align*}
    \norm{\Pi_{\suc} \bsampB \ket{\varphi}}^2
     & = \sum_{x, u, w, (f(x'))_{x' \notin w_{\vec{x}}}} \norm[\bigg]{\sum_{(f(x'))_{x' \in w_{\vec{x}}}} \alpha_{x,u,w,f} \Pi_{\suc} \bsampB  \ket{x,u,w} \ket{f}}^2 \\
     & \leq \sum_{x, u, w, (f(x'))_{x' \notin w_{\vec{x}}}} \pt[\bigg]{\sum_{(f(x'))_{x' \in w_{\vec{x}}}} \abs{\alpha_{x,u,w,f}}^2} \pt[\bigg]{\sum_{(f(x'))_{x' \in w_{\vec{x}}}} \norm{\Pi_{\suc} \bsampB \ket{x,u,w} \ket{f}}^2} \\
     & \leq \norm{\ket{\varphi}}^2 \cdot 3^K p^{K-k}.
  \end{align*}
\end{proof}

We can now conclude the proof of the main result.

\begin{theorem}
  \label{Thm:QSearch}
  Let $p \in [0,1]$. The success probability of finding $K$ marked items in a random function $f : [M] \ra \rn$ where $f(x) = 1$ with probability $p$ for each $x$ is at most~$\bo{p(T/K)^2}^K$ for any algorithm making $T \geq K$ quantum queries to $f$.
\end{theorem}

\begin{proof}
    Let $\ket{\psi_T}$ (resp. $\ket{\phi_T}$) denote the state of the algorithm after $T$ queries in the standard (resp. recording) query model. According to Theorem~\ref{Thm:RecModel}, we have $\ket{\psi_T} = \bsampB \ket{\phi_T}$. Thus, by using the fact that $\id = \sum_{k=0}^{K-1} \Pi_{= k} + \Pi_{\geq K}$, the success probability $\sigma = \norm{\Pi_{\suc} \ket{\psi_T}}^2$ satisfies
    \begin{align*}
     \sqrt{\sigma}
      & \leq \sum_{k = 0}^{K-1} \norm{\Pi_{\suc} \bsampB \Pi_{= k} \ket{\phi_T}} + \norm{\Pi_{\suc} \bsampB \Pi_{\geq K} \ket{\phi_T}} \tag*{by the triangle inequality} \\
      & \leq \sum_{k = 0}^{K-1} 3^{K/2} p^{(K-k)/2} \norm{\Pi_{= k} \ket{\phi_T}} + \norm{\Pi_{\geq K} \ket{\phi_T}} \tag*{by Proposition~\ref{Prop:QSearch-Succ}}\\
      & \leq \sum_{k = 0}^{K} 3^{K/2} p^{(K-k)/2} \binom{T}{k} (4\sqrt{p})^k \tag*{by Proposition~\ref{Prop:QSearch-Prog}} \\
      & \leq \pt*{\frac{4e\sqrt{3p}T}{K}}^K \tag*{since $\sum_{k = 0}^{K} \binom{T}{k} \leq (eT/K)^K$.}
    \end{align*}
\end{proof}

%% file: Tradeoff.tex
\subsection{Time-space tradeoff for Collision Pairs Finding}
\label{Sec:Coll-TS}

We use the time lower bound obtained in Section~\ref{Sec:Coll} to derive a time-space tradeoff for the problem of finding $K$ disjoint collisions in a random function $f : [M] \ra [N]$. We recall that the output is produced in an online fashion (Section~\ref{Sec:ModelSpace}), meaning that a collision can be output as soon as it is discovered. The length of the output is not counted towards the space bound. We allow the same collision to be output several times, but it contributes only once to the total count.

\begin{theorem}
  \label{Thm:TS-DisColl}
  Any quantum algorithm for finding $K$ disjoint collisions in a random function $f : [M] \ra [N]$ with success probability $2/3$ must satisfy a time-space tradeoff of~$T^3 S \geq \om{K^3 N}$.
\end{theorem}

\begin{proof}
  We necessarily have $T \geq \om{K^{2/3}N^{1/3}}$ by the time lower bound proved in Theorem~\ref{Thm:FindDisColl}. If $S \geq K$ then it readily implies that $T^3 S \geq \om{K^3 N}$. Thus, in the rest of the proof we can also assume that $S < K$. Our approach relies on the time-segmentation method for large-output problems, which is used for instance in \cite{BFK+81j,KSW07j}. Fix any quantum circuit~$\ci$ in the space-bounded model of Section~\ref{Sec:ModelSpace} running in time $T$ and using $S > \om{\log N}$ qubits of memory. The circuit~$\ci$ is partitioned into $L = T/T'$ consecutive sub-circuits $\ci_1 \conc \ci_2 \conc \dots \conc \ci_L$ each running in time $T' = c S^{2/3}N^{1/3}$ (for some small enough constant~$c$), where the circuit~$\ci_j$ takes as input the output memory of $\ci_{j-1}$ for each $j \in [L]$. Define~$X_j$ to be the random variable that counts the number of (mutually) disjoint collisions that $\ci$ outputs between time $(j-1) T'$ and $j T'$ (i.e. in the sub-circuit~$\ci_j$) when the input is a random function $f : [M] \ra [N]$. More precisely,~$X_j$ is the increment in the number of disjoint collisions observed by measuring the output register at time $(j-1) T'$ and then at time $j T'$. The algorithm must satisfy $\sum_{j=1}^L \ex{X_j} \geq \om{K}$ to be correct. We claim that it outputs at most $3S$ collisions in expectation in each segment of the computation. Assume towards a contradiction that $\ex{X_j} \geq 3S$ for some $j$. Since~$X_j$ is bounded between $0$ and $N$ we have $\pr{X_j > 2S} \geq S/N$. Consequently, by running~$\ci_j$ on the completely mixed state on $S$ qubits we obtain $2S$ disjoint collisions with probability at least~$S/N \cdot 2^{-S}$ in time $T'$ (this is akin to a union-bound argument). However, by Theorem~\ref{Thm:FindDisColl}, no quantum algorithm can find more than~$2S$ disjoint collisions in time $T' = c S^{2/3}N^{1/3}$ with success probability larger than $4^{-S}$ (when~$c$ is small enough). This contradiction implies that~$\ex{X_j} \leq 3S$ for all $j$. Consequently, there must be at least $L \geq \om{K/S}$ sub-circuits in order to have $\sum_{j=1}^L \ex{X_j} \geq \om{K}$. Since each sub-circuit runs in time $c S^{2/3}N^{1/3}$ the overall running time of~$\ci$ is $T \geq \om{L \cdot S^{2/3}N^{1/3}} \geq \om{K N^{1/3}/S^{1/3}}$.
\end{proof}

As an illustration of the above result, we obtain that any quantum algorithm for finding $\ta{N}$ disjoint collisions in a random function must satisfy a time-space tradeoff of $T S^{1/3} \geq \om{N^{4/3}}$. We prove that any improvement to this lower bound would imply a breakthrough for the Element Distinctness problem.

\begin{problem}
  The Element Distinctness problem $\ed_N$ on domain size $N$ consists of finding a collision in a random function $f : [N] \ra [N^2]$.
\end{problem}

It is well-known that the query complexity of Element Distinctness is $T = \ta{N^{2/3}}$ \cite{AS04j,Amb07j}. However, it is a long-standing open problem to find any quantum time-space lower bound (even classically the question is not completely settled yet \cite{Yao94j,BSSV03j}). Here, we show that \emph{any} improvement to Theorem~\ref{Thm:TS-DisColl} would imply a non-trivial time-space tradeoff for Element Distinctness. This result relies on a reduction presented in Algorithm~\ref{Algo:reduction} and analyzed in Proposition~\ref{Prop:reduction} (the constants $c_0$, $c_1$, $c_2$ are chosen in the proof).

\algobox{Algo:reduction}{Finding collisions by using $\ed_{\sqrt{N}}$.}{
{\bf Input:} a function $f : [N] \ra [N]$ containing at least $c_0 N$ collisions. \\
{\bf Output:} at least $c_1 N$ collisions in $f$ (not necessarily disjoint).

\begin{enumerate}[leftmargin=*]
  \item Repeat $c_2 N$ times:
    \begin{enumerate}
      \item Sample a 4-wise independent hash function $h : [\sqrt{N}] \ra [N]$ and store it in memory.
      \item Run an algorithm for $\ed_{\sqrt{N}}$ on input $f \circ h : [\sqrt{N}] \ra [N]$. If it finds a collision $(f \circ h (i), f \circ h (j))$ check if $h(i) \neq h(j)$ and output the collision $(h(i),h(j))$ in this case.
    \end{enumerate}
\end{enumerate}
}

\begin{proposition}
  \label{Prop:reduction}
  Let $N$ be a square number. If there is an algorithm solving $\ed_N$ in time $T_N$ and space~$S_N$ then Algorithm~\ref{Algo:reduction} runs in time $\bo[\bi]{N T_{\sqrt{N}}}$ and space $\bo[\big]{S_{\sqrt{N}}}$, and it finds $c_1 N$ collisions in any function $f : [N] \ra [N]$ containing at least $c_0 N$ collisions.
\end{proposition}

\begin{proof}
  The space complexity is fulfilled by using any space-efficient construction of 4-wise independent hash functions. For the time complexity, we choose $c_0 = 150$, $c_1 = 1/10^4$ and $c_2 = 8$. We study the probabilities of the following events to occur in a fixed round of Algorithm~\ref{Algo:reduction}:
  \begin{itemize}
    \item \textbf{Event A:} The function $h$ is collision free (i.e. $h(i) \neq h(j)$ for all $i \neq j$).
    \item \textbf{Event B:} The image of $h$ does not contain any pair of values output in a previous round.
    \item \textbf{Event C:} The function $f \circ h : [\sqrt{N}] \ra [N]$ contains a collision.
    \item \textbf{Event D:} The algorithm for $\ed_{\sqrt{N}}$ finds a collision at step 2.b.
  \end{itemize}
  Algorithm~\ref{Algo:reduction} succeeds if the event $A \wedge B \wedge C \wedge D$ occurs during at least $c_1 N$ rounds. We now lower bound the probability of this event happening.

  For \textbf{event A}, consider the random variable $X = \sum_{i \neq j \in [\sqrt{N}]} 1_{h(i) = h(j)}$. Using that~$h$ is pairwise independent, we have $\ex{X} = \binom{\sqrt{N}}{2} \frac{1}{N} \leq \frac{1}{2}$. By Markov's inequality, $\pr{ A} = 1 - \pr{X \geq 1} \geq \frac{1}{2}$.

  For \textbf{event B}, let us assume that $k <  c_1 N$ collisions $(x_1,x_2),\dots,(x_{2k-1},x_{2k})$ have been output so far. For any $i \in [k]$, the probability that both $x_{2i-1}$ and $x_{2i}$ belong to $\set{h(1), \dots, h(\sqrt{N})}$ is at most $\binom{\sqrt{N}}{2} \frac{2}{N^2} \leq \frac{1}{N}$ since $h$ is pairwise independent. By a union bound, $\pr{ B} \geq 1 - \frac{k}{N} \geq 1 - c_1$.

  For \textbf{event C}, let us consider the binary random variables $Y_{i,j} = 1_{f \circ h (i) = f \circ h (j)}$ for $i \neq j \in [\sqrt{N}]$, and let $Y = \sum_{i \neq j} Y_{i,j}$ be twice the number of collisions in $f \circ h$. Note that we may have $Y_{i,j} = 1$ because $h(i) = h(j)$ (this is taken care of in event A). For each $y \in [N]$, let $N_y = \abs{\set{x : f(x) = y}}$ denote the number of elements that are mapped to $y$ by $f$. Using that $h$ is 4-wise independent, for any $i \neq j \neq k \neq \ell$ we have,
    \[
    \left\{
        \begin{array}{l}
            \pr{Y_{i,j} = 1} = \frac{1}{N^2} \sum_{y \in [N]} N_y^2, \\ [6pt]
            \pr{Y_{i,j} = 1 \wedge Y_{i,k} = 1} =  \frac{1}{N^3} \sum_{y \in [N]} N_y^3, \\ [10pt]
            \pr{Y_{i,j} = 1 \wedge Y_{k,\ell} = 1} = \pr{Y_{i,j} = 1} \cdot \pr{Y_{k,\ell} = 1}.
        \end{array}
    \right.
    \]
   Thus, $\ex{Y} = \binom{\sqrt{N}}{2} \frac{1}{N^2} \sum_{y \in [N]} N_y^2$ and
    $
      \var{Y}  = \sum_{\set{i,j}} \var{Y_{i,j}} + \sum_{\set{i,j} \neq \set{i,k}} \mathrm{Cov}\bc{Y_{i,j},Y_{i,k}} + \sum_{\set{i,j} \cap \set{k,\ell} = \varnothing} \mathrm{Cov}\bc{Y_{i,j},Y_{k,\ell}}
               \leq \sum_{\set{i,j}} \ex{Y_{i,j}^2} + \sum_{\set{i,j} \neq \set{i,k}} \ex{Y_{i,j}Y_{i,k}}
               = \ex{Y} + 6\binom{\sqrt{N}}{3} \frac{1}{N^3} \sum_{y} N_y^3
    $
   where we used that $Y_{i,j}$ and $Y_{k,\ell}$ are independent when $i \neq j \neq k \neq \ell$. The term $\sum_{y \in [N]} N_y^2$ is equal to the number of pairs $(x,x') \in [N]^2$ such that $f(x) = f(x')$. Each collision in $f$ gives two such pairs, and we must also count the pairs $(x,x)$. Thus, $\sum_{y \in [N]} N_y^2 \geq (1+2c_0)N$. Moreover, $\sum_{y \in [N]} N_y^3 \leq (\sum_{y \in [N]} N_y^2)^{3/2}$. Consequently,
    \[
      \frac{\var{Y}}{\ex{Y}^2}
      \leq \frac{1+2\sqrt{N} \pt[\big]{\frac{1}{N^2} \sum_{y \in [N]} N_y^2}^{1/2}}{\binom{\sqrt{N}}{2} \frac{1}{N^2} \sum_{y \in [N]} N_y^2}
      \leq \frac{N}{\binom{\sqrt{N}}{2} (1+2c_0)} + \frac{2N}{\binom{\sqrt{N}}{2} \sqrt{1+2c_0}}
        \leq \frac{4 + 8\sqrt{1+2c_0}}{1+2c_0}.
    \]
  Finally, according to Chebyshev's inequality, $\pr{Y = 0} \leq \pr{\abs{Y - \ex{Y}} \geq \ex{Y}} \leq \frac{\var{Y}}{\ex{Y}^2}$. Thus, $\pr{ C} = 1 - \pr{Y = 0} \geq 1 - \frac{4 + 8\sqrt{1+2c_0}}{1+2c_0}$.

  For \textbf{event D}, we have $\pr{ D \given A \wedge B \wedge C} \geq 2/3$ assuming the algorithm for solving $\ed_{\sqrt{N}}$ succeeds with probability $2/3$.

  The probability of the four events happening together is $\pr{ A \wedge B \wedge C \wedge D} = \pr{ D \given A \wedge B \wedge C} \cdot \pr{ A \wedge B \wedge C} \geq \pr{ D \given A \wedge B \wedge C} \cdot (\pr{ A} + \pr{ B} + \pr{ C} - 2) \geq \frac{2}{3} \cdot \pt*{\frac{1}{2} - c_1 - \frac{4 + 8\sqrt{1+2c_0}}{1+2c_0}} \geq 1/250$. Let $\tau$ be the number of rounds after which $c_1 N$ collisions have been found (i.e. $A \wedge B \wedge C \wedge D$ has occurred $c_1 N$ times). We have $\ex{\tau} \leq 250 c_1 N$ and by Markov's inequality $\pr{\tau \geq c_2 N} \leq 250 c_1/c_2 \leq 1/3$. Thus, with probability at least $2/3$, Algorithm~\ref{Algo:reduction} outputs at least~$c_1 N$ collisions in $f$.
\end{proof}

We now use the above reduction to transform any low-space algorithm for Element Distinctness into one for finding $\om{N / \log N}$ disjoint collisions in a random function. Observe that Algorithm~\ref{Algo:reduction} does not necessarily output collisions that are mutually disjoint. Nevertheless, there is a small probability that a random function $f : [M] \ra [N]$ contains multi-collisions of size larger than~$\log N$ when $M \approx N$ \cite{FO89c}. Thus, there is only a $\log N$ loss in the analysis.

\begin{proposition}
  \label{Prop:TS-ED}
  Suppose that there exists a quantum algorithm for solving Element Distinctness on domain size $N$ that satisfies a time-space tradeoff of $T^{\alpha} S^{\beta} \leq \wbo[\big]{N^{2(\gamma - \alpha)}}$ for some constants $\alpha, \beta, \gamma$. Then, there exists a quantum algorithm for finding $\om{N / \log N}$ disjoint collisions in a random function $f : [10 N] \ra [N]$ that satisfies a time-space tradeoff of $T^{\alpha} S^{\beta} \leq \wbo{N^{\gamma}}$.
\end{proposition}

\begin{proof}
  We use the constants $c_0, c_1, c_2$ specified in the proof of Proposition~\ref{Prop:reduction}. First, we note that a random function $f: [10N] \ra [N]$ contains $c_0 N$ collisions and no multi-collisions of size larger than~$\log(N)$ with large probability \cite{FO89c}. Consequently, any set of $c_1 N$ collisions contains at least~$c_1 N / \log N$  disjoint collisions with large probability. Assume now that there exists an algorithm solving $\ed_{\sqrt{10N}}$ in time $T_{\sqrt{10N}}$ and space $S_{\sqrt{10N}}$ such that $\pt[\big]{T_{\sqrt{10N}}}^{\alpha} S_{\sqrt{10N}}^{\beta} \leq \wbo{N^{\gamma-\alpha}}$. Then, by plugging it into Algorithm~\ref{Algo:reduction}, one can find $c_1 N / \log N$ disjoint collisions in a random function $f: [10N] \ra [N]$ in time $T = \bo[\big]{N T_{\sqrt{10N}}}$ and space~$S = \bo[\big]{S_{\sqrt{10N}}}$. We derive from the above tradeoff that $T^{\alpha} S^{\beta} \leq \wbo{N^{\gamma}}$.
\end{proof}

As an application of Proposition~\ref{Prop:TS-ED}, we obtain the following result regarding the hardness of finding $\wom{N}$ collisions.

\begin{corollary}
  \label{Cor:TS-ED-All}
  Suppose that there exists $\eps \in (0,1)$ such that any quantum algorithm for finding~$\wom{N}$ disjoint collisions in a random function $f : [10N] \ra [N]$ must satisfy a time-space tradeoff of $T S^{1/3} \geq \wom{N^{4/3 + \eps}}$. Then, any quantum algorithm for solving Element Distinctness on domain size $N$ must satisfy a time-space tradeoff of $T S^{1/3} \geq \wom{N^{2/3 + 2\eps}}$.
\end{corollary}

We conjecture that the optimal tradeoff for finding $K$ collisions is $T^2 S = \ta{K^2 N}$, which would imply an optimal time-space tradeoff of $T^2 S \geq \wom{N^2}$ for Element Distinctness.

\begin{conjecture}
  \label{Conj:Coll}
  Any quantum algorithm for finding $K$ disjoint collisions in a random function $f : [M] \ra [N]$ with success probability $2/3$ must satisfy a time-space tradeoff of~$T^2 S \geq \om{K^2 N}$.
\end{conjecture}

\begin{corollary}
  \label{Cor:ED}
  If Conjecture~\ref{Conj:Coll} is true, then any quantum algorithm for solving Element Distinctness with success probability $2/3$ must satisfy a time-space tradeoff of~$T^2 S \geq \wom{N^2}$.
\end{corollary}

Finally, we describe a quantum algorithm that achieves the tradeoff of $T^2 S \leq \wbo{K^2 N}$. In order to simplify the analysis, we do not require the collisions to be disjoint.

\algobox{Algo:BHTvariant}{Finding $K$ collision pairs in $f : [N] \ra [N]$ using a memory of size $S$.}{
\begin{enumerate}[leftmargin=*]
  \item Repeat $\bo{K/S}$ times:
    \begin{enumerate}
      \item Sample a subset $G \subset [N]$ of size $S$ uniformly at random.
      \item Construct a table containing all pairs $(x,f(x))$ for $x \in G$. Sort the table according to the second entry of each pair.
      \item Define the function $g : [N] \setminus G \ra \rn$ where $g(x) = 1$ iff there exists $x' \in G$ such that $f(x) = f(x')$. Run the Grover search algorithm \cite{BBHT98j} on~$g$, by using the table computed at step 1.b, to find all pairs $(x,x') \in G \times ([N] \setminus G)$ such that $f(x) = f(x')$. Output all of these pairs.
    \end{enumerate}
\end{enumerate}
}

\begin{proposition}
  \label{Prop:BHTvariant}
  For any $1 \leq K \leq \bo{N}$ and $\wom{\log N} \leq S \leq \wbo{K^{2/3} N^{1/3}}$, there exists a quantum algorithm that can find $K$ collisions in a random function $f : [N] \ra [N]$ with probability at least $2/3$ by making $T = \wbo{K \sqrt{N/S}}$ queries and using $S$ qubits of memory.
\end{proposition}

\begin{proof}
  We prove that Algorithm~\ref{Algo:BHTvariant} satisfies the statement of the proposition. For simplicity, we do not try to tune the hidden factors in the big O notations.

  The probability that a fixed pair $(x,x')$ satisfies $(x,x') \in G \times ([N] \setminus G)$ for at least one iteration of step 1 is $\om{K/S \cdot S/N \cdot (1-S/N)} = \om{K/N}$. Since a random function $f : [N] \ra [N]$ contains $\om{N}$ collisions with high probability, the algorithm encounters $\om{K}$ collisions in total. Thus, if the Grover search algorithm never fails we obtain the desired number of collisions.

  The expected number of pre-images of $1$ under $g$ is $\bo{S}$. Consequently, the complexity of Grover's search at step 1.c is $\bo{\sqrt{SN}}$. The overall query complexity is $T = \wbo{K/S \cdot \sqrt{SN}} = \wbo{K \sqrt{N/S}}$, and the space complexity is $\wbo{S}$.
\end{proof}


\subsection{Time-space tradeoff for Sorting}
\label{Sec:Sort-TS}

We use the time lower bound obtained in Section~\ref{Sec:QSearch} to reprove the time-space tradeoff for the Sorting problem described in~\cite[Theorem 21]{KSW07j}. The input to the Sorting problem is represented as a function $f : [N] \ra \set{0,1,2}$ (we do not need to consider a larger range for the proof). The objective is to output \emph{in order} a sequence $x_1,\dots,x_N \in [N]$ of distinct integers such that $f(x_1) \geq f(x_2) \geq \dots \geq f(x_N)$ with probability at least $2/3$.

We weaken the space-bounded model described in Section~\ref{Sec:ModelSpace} by removing the flag register that allowed the algorithm to choose when to update the output register. As a consequence, each of the $N$ elements of the sorted sequence must be written on the output register at a predetermined time of the computation. Note that, without this condition, it is easy to sort a function $f : [N] \ra \set{0,1,2}$ in time $T = \bo{N}$ and space $S = \bo{\log N}$. The same limitation was present in previous work~\cite{KSW07j} and it is an open problem to get rid of it. It will require considering input functions with a larger range, as is the case in classical tradeoffs (e.g.~\cite{Bea91j}).

\begin{theorem}
  \label{Thm:TS-DisSort}
  Any quantum algorithm for sorting any function $f : [N] \ra \set{0,1,2}$ with success probability $2/3$ must satisfy a time-space tradeoff of $T^2 S \geq \om{N^3}$.
\end{theorem}

\begin{proof}
  The proof is a modified version of~\cite[Theorem 21]{KSW07j} adapted to our version of the $K$-Search problem and to our slightly more general computational model. Given a circuit $\ci$ that runs in time $T$ and space $S \geq \om{\log N}$, we partition it into $L = T/T'$ consecutive sub-circuits $\ci_1 \conc \ci_2 \conc \dots \conc \ci_L$ each running in time $T' = c\sqrt{SN}$ (for some small enough constant $c$). Assume towards a contradiction that a circuit~$\ci_j$ outputs the elements of ranks $r, r+1, \dots, r+2S-1$ for some $r \leq N/2$. We use $\ci_j$ to solve the $K$-search problem for $K = 2S$ as follows. Given an input $g : [N/2] \ra \rn$ to the $K$-search problem where $g(x) = 1$ with probability $p = \frac{6S}{N}$ for each $x$, define the function $f : [N] \ra \set{0,1,2}$ where
    \[f(x) =
    \left\{
        \def\arraystretch{1.2}
        \begin{array}{ll}
            2         & \text{if $x < r$,} \\
            g(x-r+1) & \text{if $r \leq x < r + N/2$,} \\
            0         & \text{if $x \geq r + N/2$.}
        \end{array}
    \right.\]
  Note that the function $g$ contains at least $2S$ marked items with probability at least $2S/N$. Thus, if the circuit $\ci$ is run on the input $f$, then the indices output by the sub-circuit~$\ci_j$ contain the position of $2S$ marked items with probability at least $2/3 \cdot 2S/N$. Consequently, by running~$\ci_j$ on the completely mixed state on $S$ qubits we can find $2S$ marked items under~$g$ with probability at least $2/3 \cdot 2S/N \cdot 2^{-S}$ in time $T'$. However, by Theorem~\ref{Thm:QSearch}, any such algorithm must succeed with probability at most $4^{-S}$ (when $c$ is small enough). This contradiction implies that there must be at least $L \geq \om{N/S}$ sub-circuits in $\ci$. Thus, the running time of $\ci$ is $T \geq \om{L \cdot \sqrt{SN}} \geq \om{N^{3/2}/\sqrt{S}}$.
\end{proof}

The time-space tradeoffs for the Boolean matrix-vector product~\cite[Theorem 23]{KSW07j} and the Boolean matrix product~\cite[Theorem 25]{KSW07j} problems can be reproved in a similar way.